\documentstyle[preprint,aps]{revtex} 
\begin{document}
\draft
\date{to appear in {\sl The Philosophical Magazine B}}
\title{Exact results for the optical absorption of strongly 
correlated electrons in a half-filled Peierls-distorted chain}
\author{F.~Gebhard\footnote{e-mail: {\tt florian\verb2@2gaston.ill.fr}}}
\address{ILL Grenoble, B.~P.\ 156x, F-38042 Grenoble Cedex 9, France}
\author{K.~Bott, M.~Scheidler, P.~Thomas, S.~W.~Koch}
\address{Dept.~of Physics and Materials Sciences Center,\\
Philipps University Marburg, D-35032~Marburg, Germany}

\maketitle

\begin{abstract}%
In this second of three articles on the optical absorption
of electrons in a half-filled Peierls-distorted chain
we present exact results for strongly correlated tight-binding electrons.
In the limit of a strong on-site interaction~$U$ we map the 
Hubbard model onto the Harris-Lange model which can be solved exactly
in one dimension 
in terms of spinless fermions for the charge excitations.
The exact solution allows for an interpretation of the charge dynamics
in terms of {\em parallel\/}
Hubbard bands with a free-electron dispersion of band-width~$W$, 
separated by the Hubbard interaction~$U$.
The spin degrees of freedom enter the expressions for
the optical absorption only via a
momentum dependent but static ground state expectation value.
The remaining spin problem can be traced out exactly 
since the eigenstates of the Harris-Lange model are
spin-degenerate.
This corresponds to the Hubbard model at temperatures large compared to
the spin exchange energy.

Explicit results are given for the optical
absorption in the presence
of a lattice distortion~$\delta$ and a nearest-neighbor interaction~$V$.
We find that the optical absorption for $V=0$
is dominated by a peak at~$\omega=U$ and broad but weak
absorption bands for $| \omega -U | \leq W$. For an appreciable
nearest-neighbor interaction, $V>W/2$, almost all spectral weight
is transferred to Simpson's exciton band which is eventually Peierls-split.

\vskip2cm\noindent
PACS1996: 71.10.Fd, 71.20.Rv, 36.20.Kd
\end{abstract}

\newpage
\section{Introduction}

Some charge-transfer salts are understood as strongly correlated
one-dimensional electron systems for half-filled bands
(Mott-Hubbard insulators)
(Farges 1994); (Alc\'{a}cer, Brau, and Farges 1994).   
The extended Hubbard model for interacting electrons
on a Peierls-distorted 
chain at half-filling is considered appropriate for these
materials (Mazumdar and Dixit 1986); 
(Fritsch and Ducasse 1991); (Mila 1995).  
There are only few studies of the optical, i.~e., finite
frequency properties
of correlated electron systems
since their calculation
is a formidable task
(Kohn 1964); (Maldague 1977); 
(Lyo and Galinar 1977); (Lyo 1978); (Galinar 1979);
(Campbell, Gammel, and Loh 1988); (Mahan 1990); 
(Shastry and Sutherland 1990);
(Stafford, Millis, and Shastry 1991); 
(Fye, Martins, Scalapino, Wagner, and Hanke 1992);
(Stafford and Millis 1993).     

In this second article on the optical absorption
of electrons in half-filled Peierls-distorted chains
we present a detailed analysis of the optical
absorption in the limit of strong correlations and for a half-filled band
where the charge and spin dynamics decouple.
We find that the physics of the half-filled Hubbard model at strong
correlations is determined by the upper and lower Hubbard band
for the charges which are {\em parallel\/} bands.
This essential feature and its significant consequences
have been missed in earlier analytical and numerical investigations.
In this work we include a finite lattice
dimerization and nearest-neighbor interaction between the electrons,
i.~e., we analyze the extended dimerized Hubbard model at
strong correlations.

The paper is organized as follows.
In section~\ref{Hamilts} we address the
Hubbard model at strong coupling from which we derive the Harris-Lange model
which determines the motion of the charge degrees of freedom.
In the ground state there are no free charges but only singly occupied
lattice sites. 
The exact spectrum and eigenstates of the Harris-Lange model
are presented in section~\ref{exactsolution} for the translational
invariant and the dimerized case.
The results can be interpreted in terms of two {\em parallel\/}
Hubbard bands for the charges which are eventually split
into Peierls subbands. Optical absorption can now formally
be treated as if we had {\em independent\/} (spinless) Fermions.

Unfortunately, this simple band structure interpretation is blurred by
the spin degrees of freedom which enter the expressions
for the optical absorption in terms
of a very complicated ground state expectation value.
In section~\ref{optabsHL} we treat the case of the Harris-Lange
model where all 
spin configurations are equally possible ground states.
This corresponds to the Hubbard model at temperatures
large compared to the spin exchange energy.
It allows the calculation of
the optical absorption even in
the presence of a Peierls distortion and a nearest-neighbor
interaction between the charges.
A summary and outlook closes our presentation.
Some details of the calculations are left to the appendices.

\section{Strongly correlated Mott-Hubbard insulators}
\label{Hamilts}

\subsection{Tight-binding electrons on a Peierls-distorted chain}
\label{Peidistrot}

For narrow-band materials
the electron transfer is limited to nearest neighbors only.
In standard notation of second quantization
the Hamiltonian for electrons in the tight-binding approximation reads
\begin{equation}
\hat{T}(\delta)= - t \sum_{l=1,\sigma}^{L}
\left(1+ (-1)^l\delta\right)
\left(
\hat{c}_{l,\sigma}^+ \hat{c}_{l+1,\sigma}^{\phantom{+}} 
+ \hat{c}_{l+1,\sigma}^+ \hat{c}_{l,\sigma}^{\phantom{+}}
\right) 
\end{equation}
where $\delta$ describes the effect of
bond-length alternation on the electron transfer amplitudes.
As usual the Hamiltonian can be diagonalized in momentum space.
We apply periodic boundary conditions, and introduce the
Fourier transformed electron operators
as $\hat{c}_{k,\sigma}^+=\sqrt{1/L} \sum_{l=1}^L
\exp(ikla) \hat{c}_{l,\sigma}^+$ for the~$L$ momenta~$k=2\pi m/(La)$,
$m=-(L/2),\ldots (L/2)-1$.
We may thus write
\begin{equation}
\hat{T}(\delta) = \sum_{|k|\leq \pi/(2a),\sigma}
\epsilon(k)\left(
\hat{c}_{k,\sigma}^+\hat{c}_{k,\sigma}^{\phantom{+}}
-
\hat{c}_{k+\pi/a,\sigma}^+\hat{c}_{k+\pi/a,\sigma}^{\phantom{+}}
\right)
-i \Delta(k)
\left( \hat{c}_{k+\pi/a,\sigma}^+\hat{c}_{k,\sigma}^{\phantom{+}}
-
\hat{c}_{k,\sigma}^+\hat{c}_{k+\pi/a,\sigma}^{\phantom{+}}\right)
\label{Tink}
\end{equation}
with the dispersion relation~$\epsilon(k)$ and hybridization
function~$\Delta(k)$ defined as
\begin{mathletters}
\begin{eqnarray}
\epsilon(k) &=& -2t\cos(ka)\label{gl3a} \\[3pt]
\Delta(k) &=& 2t\delta\sin(ka)\; .
\end{eqnarray}
\end{mathletters}%
The Hamiltonian can easily be diagonalized in $k$-space.
The result is (Gebhard, Bott, Scheidler, Thomas, and Koch I 1996)   
\begin{equation}
\hat{T}(\delta)
= \sum_{|k|\leq \pi/(2a),\sigma} E(k) (\hat{a}_{k,\sigma,+}^+
\hat{a}_{k,\sigma,+}^{\phantom{+}}
- \hat{a}_{k,\sigma,-}^+ \hat{a}_{k,\sigma,-}^{\phantom{+}})
\; .
\label{Tdiapeierls}
\end{equation}
Here, $\pm E(k)$ is the dispersion relation for the upper~($+$)
and lower~($-$) Peierls band,
\begin{equation}
E(k) = \sqrt{\epsilon(k)^2 + \Delta(k)^2} \; .
\label{peierlsen}
\end{equation}
The new Fermion quasi-particle operators $\hat{a}_{k,\sigma,\pm}^{+}$
for these two bands are related to the original electron operators by
\begin{mathletters}
\label{mixamp}
\begin{eqnarray}
\hat{a}_{k,\sigma,-}^{\phantom{+}} &=&
\alpha_k \hat{c}_{k,\sigma}^{\phantom{+}} + i \beta_k
\hat{c}_{k+\pi,\sigma}^{\phantom{+}} \\[6pt]
\hat{a}_{k,\sigma,+}^{\phantom{+}} &=&
\beta_k \hat{c}_{k,\sigma}^{\phantom{+}} -i \alpha_k
\hat{c}_{k+\pi,\sigma}^{\phantom{+}}
\end{eqnarray}
\end{mathletters}%
with
\begin{mathletters}
\label{alphabeta}
\begin{eqnarray}
\alpha_k &=& \sqrt{ \frac{1}{2} \left( 1 - \frac{\epsilon(k)}{E(k)}
\right) }\\[6pt]
\beta_k &=& - \sqrt{ \frac{1}{2} \left( 1 + \frac{\epsilon(k)}{E(k)}
\right)} {\rm sgn}\left(\Delta(k)\right)
\end{eqnarray}
\end{mathletters}%
Details of the upper transformation and the optical absorption
of this model are presented in (Gebhard {\em et al.} I 1996).    

\subsection{Hubbard model}
\label{Hubbard-Model}

The only spinful interacting electron model that can be solved exactly
for all values of the interaction strength is the
Hubbard model in one dimension
(Hubbard 1963); 
(Gebhard and Ruckenstein 1992); 
(E\ss ler and Korepin 1994);   
(Gebhard, Girndt, and Ruckenstein 1994);
(Bares and Gebhard 1995).
For narrow-band materials
the electron transfer is limited to nearest neighbors only,
and the interaction is supposed to be described by the
purely local (Hub\-bard-)interaction of strength~$U$,
\begin{eqnarray}
\hat{H}_{\rm Hubbard}&=& \hat{T} +U \hat{D}\nonumber\\[6pt]
\label{Hubb-Model}
\hat{D}&=&\sum_{l} \hat{D}_{l}
= \sum_l \hat{n}_{l,\uparrow}\hat{n}_{l,\downarrow} \; ,
\end{eqnarray}
where
$\hat{n}_{l,\sigma}=\hat{c}_{l,\sigma}^+\hat{c}_{l,\sigma}^{\phantom{+}}$
is the local density of $\sigma$-electrons, and
$\hat{T}=\hat{T}(\delta=0)$.

The model~(\ref{Hubb-Model}) poses a very difficult many-body problem.
Its spectrum and, in particular, 
its elementary excitations
can be obtained from the Bethe Ansatz solution
(Lieb and Wu 1968); (Shastry, Jha, and Singh 1985); (Andrei 1995).   
Its low-energy properties including the DC-conductivity,
$\sigma_{\rm DC}={\rm Re}\{\sigma(\omega=0)\}$,
can explicitly be obtained from the corresponding $g$-ology
Hamiltonian (Schulz 1990); (Schulz 1991) or from conformal field theory   
(Frahm and Korepin 1990); (Frahm and Korepin 1991);
(Kawakami and Yang 1990); (Kawakami and Yang 1991).    
The Hubbard model
describes a (correlated) metal for all~$U>0$ for less than half-filling.
Unfortunately, the Bethe Ansatz solution
does not allow the direct calculation of transport properties at
finite frequencies.

At half-filling the one-dimensional Hubbard model
describes a Mott-insulator which implies
that~$\sigma_{\rm DC}=0$ for all~$U>0$.
The density of states for charge excitations
displays two bands,
the upper and lower Hubbard band, separated by the Mott-Hubbard gap.
This gap is defined as the jump in the chemical
potential at half filling,
\begin{mathletters}
\begin{eqnarray}
\Delta_{\rm MH}&=&\mu^+(N=L)-\mu^-(N=L) \nonumber \\[6pt]
&=& \left[ E(N=L+1)-E(N=L)\right] - \left[ E(N=L)-E(N=L-1)\right] \quad.
\end{eqnarray}
As shown by Ovchinnicov (Ovchinnicov 1969) the Mott-Hubbard gap can be    
obtained from the Lieb-Wu solution (Lieb {\em et al.} 1968) in the form  
\begin{eqnarray}
\Delta_{\rm MH} &=& \frac{16t}{U} \int_{1}^{\infty}
\frac{dy \sqrt{y^2-1}}{\sinh(2\pi t y/U)} \\[9pt]
&=& \left\{ \begin{array}{ccr}
(2W/\pi) \sqrt{4U/W} \exp(-\pi W/(2U)) & \hbox{for} & U \ll W=4t \\[6pt]
U - W + \ln(2)W^2/(2U) + {\cal O}(W^3/U^2) & \hbox{for} & U \gg W=4t
\end{array}
\right. \quad .
\end{eqnarray}
\end{mathletters}%
It is obvious that optical absorption is only
possible if $\omega \geq \Delta^{\rm MH}$.
It is further seen that the upper and lower Hubbard band
are well separated for $U\gg W$.

One might expect that the optical absorption for large interactions, $U \gg W$,
and high temperatures, $k_{\rm B}T\gg J={\cal O}(W^2/U)$,
shows the signature of a broad band-to-band 
transition for $U-W\leq \omega\leq U+W$
(units $\hbar\equiv 1$),
similar to the Peierls insulator (Gebhard {\em et al.} I 1996).    
Such considerations seemed to be supported by
analytical (Lyo {\em et al.} 1977); (Lyo 1978); (Galinar 1979)      
and numerical calculations (Campbell, Gammel, and Loh 1989).           
Below we will calculate~$\sigma(\omega>0)$ in the limit $U\gg W$,
and show that the linear absorption is actually
dominated by a singular contribution at~$\omega=U$ because the
upper and lower Hubbard band are in fact {\em parallel\/} bands.
The situation changes for $k_{\rm B}T\ll J$ which we will consider in
(Gebhard, Bott, Scheidler, Thomas, and Koch III 1996).      

\subsection{Harris-Lange model}

In the following we will address
the limit $W/U\to 0$
where matters considerably simplify since the
charge and spin degrees of freedom completely decouple
(Ogata and Shiba 1990); (Parola and Sorella 1990).                 
For example, for less than half-filling, $N \leq L$, and $U=\infty$
the eigenenergies become those
of a Fermi gas of $N_h=L-N$ holes with dispersion~$\epsilon(k)$,
and each energy level is $2^N$-fold degenerate in the thermodynamical
limit (Beni, Pincus, and Holstein 1973); (Klein 1973);                   
(Ogata {\em et al.} 1990); (Parola {\em et al.} 1990). 

To facilitate the discussion of the strong coupling
limit we map the Hubbard model onto a problem for which the number
of double occupancies is {\em conserved}. 
For a large on-site Coulomb repulsion $W/U \to 0$
it is natural to start with a spectral decomposition
of operators into those which solely act in the upper or lower Hubbard band,
and to perturbatively eliminate those parts in $\hat{H}_{\rm Hubbard}$
which couple the two bands.
For the Hubbard model this has first been achieved by Harris and
Lange (Harris and Lange 1967); (van Dongen 1994),      
and the resulting effective Hamiltonian
to lowest order in~$W/U$ will thus be called the ``Harris-Lange'' model.
It offers several advantages,
both for analytical and numerical calculations.

To carry out the spectral decomposition
we start from the case $t=0$.
The Fermi annihilation operator can be split into a part
which destroys an electron on
a single occupied site and does not 
change the energy of the state, and another part
which destroys an electron on a double occupied site 
and thus decreases the energy by $U$,
\begin{equation}
\hat{c}_{l,\sigma} = \hat{n}_{l,-\sigma} \hat{c}_{l,\sigma} +
(1 - \hat{n}_{l,-\sigma}) \hat{c}_{l,\sigma} \quad .
\end{equation}
The corresponding creation operator can be treated accordingly.

If we now turn on the hopping of electrons ($t\neq 0$)
we may split the kinetic energy operator into
\begin{mathletters}
\label{HLM}
\begin{eqnarray}
\hat{T} &=& \hat{T}_{\rm LHB} + \hat{T}_{\rm UHB}
 + \hat{T}^+ + \hat{T}^- \\[6pt]
\hat{T}_{\rm LHB} &=& (-t) \sum_{l,\sigma} \left(1-\hat{n}_{l,-\sigma}\right)
\left(
\hat{c}_{l,\sigma}^+ \hat{c}_{l+1,\sigma}^{\phantom{+}} +
\hat{c}_{l+1,\sigma}^+ \hat{c}_{l,\sigma}^{\phantom{+}}
\right)
\left( 1-\hat{n}_{l+1,-\sigma}\right) \\[6pt]
\hat{T}_{\rm UHB} &=& (-t) \sum_{l,\sigma} \hat{n}_{l,-\sigma}
\left(
\hat{c}_{l,\sigma}^+ \hat{c}_{l+1,\sigma}^{\phantom{+}} +
\hat{c}_{l+1,\sigma}^+ \hat{c}_{l,\sigma}^{\phantom{+}}
\right)
\hat{n}_{l+1,-\sigma}\\[6pt]
\hat{T}^+ &=& (-t) \sum_{l,\sigma} \left[ 
\hat{n}_{l,-\sigma}
\hat{c}_{l,\sigma}^+ \hat{c}_{l+1,\sigma}^{\phantom{+}}
\left( 1-\hat{n}_{l+1,-\sigma} \right)
+
\hat{n}_{l+1,-\sigma}
\hat{c}_{l+1,\sigma}^+ \hat{c}_{l,\sigma}^{\phantom{+}}
\left( 1- \hat{n}_{l,-\sigma}\right) \right]
\\[6pt]
\hat{T}^- &=& \left(\hat{T}^+\right)^+ \; .
\end{eqnarray}
\end{mathletters}%
The operator~$\hat{T}_{\rm LHB}$ for the lower
Hubbard band describes the hopping of holes while
doubly occupied sites can move in the
upper Hubbard band via $\hat{T}_{\rm UHB}$.
Their number is conserved by both hopping processes.
These two bands will constitute the basis for our approach.
The operator~$\hat{T}^+$ ($\hat{T}^-$) increases (decreases)
the number of double occupancies by one.

Similar to the Foldy-Wouthuysen transformation for the Dirac
equation (Bjorken and Drell 1964)              
we apply a canonical transformation that eliminates
the operators~$\hat{T}^{\pm}$ to a given order in~$t/U$,
\begin{equation}
\hat{c}_{l,\sigma} = e^{i\hat{S}(\bar{c})} \bar{c}_{l,\sigma}
e^{-i\hat{S}(\bar{c})}
\end{equation}
with $\left(\hat{S}(\bar{c})\right)^+=\hat{S}(\bar{c})$. 
As shown by Harris and Lange 
(Harris {\em et al.} 1967); (van Dongen 1994) the operator     
to lowest order in~$t/U$ reads
\begin{equation}
\hat{S}(\bar{c}) = \frac{it}{U}
\left( \hat{T}^{\bar{c},+} -\hat{T}^{\bar{c},-} \right)
\end{equation}
which can easily be verified since $\left[\hat{D},\hat{T}^{\pm}\right]_{-}=
\pm \hat{T}^{\pm}$.

The transformed Hamilton operator in the new Fermions becomes
the Harris-Lange model
\begin{equation}
\hat{H}_{\rm HL}^{\bar{c}} =
\hat{T}_{\rm LHB}^{\bar{c}} + \hat{T}_{\rm UHB}^{\bar{c}}
+ U \hat{D}^{\bar{c}} \quad ,
\end{equation}
if we neglect all correction terms to order~$t/U$ and higher.
The energies obtained from the Harris-Lange model thus agree with
those of the Hubbard model to order~$t (t/U)^{-1}$ and~$t (t/U)^0$.
For all other physical operators which do not contain a factor of~$U/t$
we may replace
\begin{equation}
\hat{c}_{l,\sigma} = \bar{c}_{l,\sigma} 
\quad .
\end{equation}
because the error is only of order $\left(t/U\right)$.
In the following we will thus make no distinction between the operators
$\hat{c}_{l,\sigma}$ and $\bar{c}_{l,\sigma}$ to lowest order in~$t/U$.

The Hamiltonian has the following symmetry. The particle-hole
transformation
\begin{mathletters}
\label{parthole}
\begin{eqnarray}
{\cal T}_{\rm ph} \hat{c}_{l,\sigma}^+ {\cal T}_{\rm ph}^{-1}& =
& i \lambda_{\sigma}
e^{i\pi l} \hat{c}_{l,-\sigma}^{\phantom{+}}
\\[6pt]
{\cal T}_{\rm ph} \hat{c}_{k,\sigma}^+ {\cal T}_{\rm ph}^{-1}& =
& i\lambda_{\sigma}
\hat{c}_{\pi/a-k,-\sigma}^{\phantom{+}}
\label{watchout}
\end{eqnarray}
with $\lambda_{\uparrow}=-\lambda_{\downarrow}=1$
is generated with the help of
\begin{eqnarray}
{\cal T}_{\rm ph} &=& e^{i\pi/2 (\hat{C}^+ +\hat{C}^-)}=
\prod_l \left[ 1-(\hat{D}_l+\hat{H}_l)
+  i(\hat{C}_l^+ +\hat{C}_l^-)\right]
\\[6pt]
\hat{C}^+&=&\left(\hat{C}^-\right)^+=\sum_l \hat{C}_l^+=
\sum_l (-1)^l \hat{c}_{l,\uparrow}^+ \hat{c}_{l,\downarrow}^+ \quad
\quad;\quad
\hat{H}_l=(1-\hat{n}_{l,\uparrow})(1-\hat{n}_{l,\downarrow})
\; .
\label{defofthecs}
\end{eqnarray}
\end{mathletters}%
The additional phase factors $i\lambda_\sigma$ are irrelevant global phases,
and can be ignored since there is always an equal number
of Fermion creation and annihilation operators of each spin species.
The operators for the motion of holes and double occupancies
are mapped into each other,
\begin{equation}
\hat{T}_{\rm UHB} \mapsto \hat{T}_{\rm LHB} \qquad
\hat{T}_{\rm LHB} \mapsto \hat{T}_{\rm UHB} \; .
\end{equation}
Furthermore,
$\left[ \hat{T}_{\rm UHB}+\hat{T}_{\rm LHB}, \hat{C}^{\pm}\right]_-=0$.
This symmetry allows for an exact solution of the model
since there is essentially
no difference in the motion of double occupancies
in the upper Hubbard band and holes in the lower Hubbard band.

The discussion above is readily generalized to the case of
dimerization in the Harris-Lange model.
The model Hamiltonian reads
\begin{equation}
\hat{H}_{\rm HL}^{\rm dim} = \hat{T}_{\rm LHB}(\delta) +
 \hat{T}_{\rm UHB}(\delta) + U\hat{D}
\end{equation}
in an obvious generalization of the kinetic operators for the
upper and lower Hubbard bands.

\subsection{Optical absorption and optical conductivity}

The dielectric function~$\widetilde{\epsilon}(\omega)$
and the coefficient for the linear optical
absorption~$\widetilde{\alpha}(\omega)$ are
given by (Haug and Koch 1990)       
\begin{mathletters}
\begin{eqnarray}
\widetilde{\epsilon}(\omega) &=& 1 +\frac{4\pi i \sigma(\omega)}{\omega}
\label{epssigma}\\[6pt]
\widetilde{\alpha}(\omega) &=&
\frac{4\pi {\rm Re}\{\sigma(\omega)\}}{n_b c}
\end{eqnarray}
\end{mathletters}%
where ${\rm Re}\{\ldots\}$ denotes the real part and
$n_b$ is the background refractive index. 
It is supposed to be frequency independent near a resonance.
Hence, the real part of the optical conductivity
directly gives the absorption spectrum of the system.

The standard result (Maldague 1977); (Mahan 1990)     
for the real part of the optical
conductivity in terms of the current-current correlation 
function~$\chi(\omega)$ is
\begin{eqnarray}
{\rm Re}\{ \sigma(\omega) \} &=&\frac{{\rm Im}\{\chi(\omega)\}}{\omega}
\\[6pt]
\chi(\omega) & =& \frac{{\cal N}_{\perp}}{La}
i \int_0^{\infty} dt e^{i\omega t} \langle
\left[\hat{\jmath}(t),\hat{\jmath}\right]_- \rangle
\end{eqnarray}
where~${\cal N}_{\perp}$ is the number of chains per unit area
perpendicular to the chain direction.

The current-current correlation function can be spectrally
decomposed in terms of exact eigenstates of the system as
\begin{equation}
\chi(\omega) = \frac{{\cal N}_{\perp}}{La}
\sum_n |\langle 0 | \hat{\jmath} | n\rangle|^2
\left[ \frac{1}{\omega +(E_n-E_0) +i\gamma} -
\frac{1}{\omega -(E_n-E_0) +i\gamma} \right] \; .
\label{decomp}
\end{equation}
Here, $|0\rangle$ is the exact ground state (energy $E_0$),
$|n\rangle$ are exact excited states (energy $E_n$),
and $\left|\langle 0 | \hat{\jmath} | n\rangle\right|^2$
are the oscillator strengths for optical transitions between them.
Although $\gamma =0^+$ is infinitesimal we may introduce $\gamma>0$
as a phenomenological broadening
of the resonances at $\omega = \pm(E_n-E_0)$.
The spectral decomposition of the real part of the optical conductivity
reads
\begin{equation}
{\rm Re}\{ \sigma(\omega) \} = \frac{{\cal N}_{\perp} \pi}{La \omega}
\sum_n\left| \langle 0 | \hat{\jmath} | n\rangle\right|^2
\left[ \delta\left( \omega
-(E_n-E_0)\right) -\delta\left( \omega +(E_n-E_0)\right) \right]
\label{speccomp}
\end{equation}
which is positive for all~$\omega$.

In the following we will always plot the dimensionless
reduced optical conductivity
\begin{equation}
\sigma_{\rm red}(\omega >0) =
\frac{\omega {\rm Re}\{\sigma(\omega>0)\}  }%
{{\cal N}_{\perp}a e^2 W  } \; .
\label{sigmared}
\end{equation}
Furthermore we replace the energy conservation~$\delta(x)$
by the smeared function
\begin{equation}
\widetilde{\delta}(x) = \frac{\gamma}{\pi(x^2+\gamma^2)} 
\end{equation}
to include effects of phonons, and experimental
resolution. For all figures we graphically checked that the sum rules of 
appendix~\ref{appsumrule} were fulfilled.

\subsection{Current operator}

As derived in (Gebhard {\em et al.} I 1996)      
the current operator is given by
\begin{equation}
\hat{\jmath} = -e \sum_{l,\sigma}
ita \left(1+ (-1)^l\delta\right)\left(1+ (-1)^l\eta\right)
\left(
\hat{c}_{l+1,\sigma}^+ \hat{c}_{l,\sigma}^{\phantom{+}} 
- \hat{c}_{l,\sigma}^+ \hat{c}_{l+1,\sigma}^{\phantom{+}}
\right)
\label{currenta}
\end{equation}
where $\eta=-|R_{l+1}-R_l-a|/a<0$ is the relative change of lattice distances
due to the Peierls distortion.
Note that $\delta$ and $\eta$ always have opposite sign.
The current operator can be split into two parts,
$\hat{\jmath}= \hat{\jmath}_{\rm intra}^{\rm H}
+\hat{\jmath}_{\rm inter}^{\rm H}$,
where $\hat{\jmath}_{\rm intra}^{\rm H}$ moves electrons between
neighboring sites without changing the
number of double occupancies or holes.
This (Hubbard-)intraband current does not change the
number of double occupancies. Hence it can be ignored
for the optical absorption in the Harris-Lange model
at half-filling.
The current operator between the two Hubbard bands
$\hat{\jmath}_{\rm inter}^{\rm H}$ can be written as
\begin{mathletters}
\begin{eqnarray}
\hat{\jmath}_{\rm inter}^{\rm H} &=&  \hat{\jmath}_{{\rm inter},+}^{\rm H}
+ \hat{\jmath}_{{\rm inter},-}^{\rm H} \\[6pt]
\hat{\jmath}_{{\rm inter},+}^{\rm H} &=&
-(itea) \sum_{l,\sigma}\left(1+(-1)^l \delta\right) \left(1+(-1)^l \eta\right)
\nonumber \\[3pt]
&& \phantom{-(itea)\sum}
\left[
\hat{n}_{l+1,-\sigma} \hat{c}_{l+1,\sigma}^+ \hat{c}_{l,\sigma}^{\phantom{+}} 
\left( 1- \hat{n}_{l,-\sigma} \right)
-
\hat{n}_{l,-\sigma} \hat{c}_{l,\sigma}^+ \hat{c}_{l+1,\sigma}^{\phantom{+}} 
\left( 1- \hat{n}_{l+1,-\sigma} \right)
\right] 
\\[6pt]
\hat{\jmath}_{{\rm inter},-}^{\rm H} &=&
-(itea) \sum_{l,\sigma}\left(1+(-1)^l \delta\right) \left(1+(-1)^l \eta\right)
\nonumber \\[3pt]
&& \phantom{-(itea)\sum}
\left[
\left(1- \hat{n}_{l+1,-\sigma}\right)
 \hat{c}_{l+1,\sigma}^+ \hat{c}_{l,\sigma}^{\phantom{+}} 
\hat{n}_{l,-\sigma} 
-
\left(1 -\hat{n}_{l,-\sigma} \right)
\hat{c}_{l,\sigma}^+ \hat{c}_{l+1,\sigma}^{\phantom{+}} 
\hat{n}_{l+1,-\sigma}
\right] 
\end{eqnarray}
\end{mathletters}%
where~$\hat{\jmath}_{\rm inter,\pm}^{\rm H}$ create and destroy
a neighboring pair of double occupancy and hole, respectively.

Next we study the action of $\hat{\jmath}_{{\rm inter},+}^{\rm H}$
on a pair of neighboring spins in a state~$|\Psi\rangle$ in
position space. It is a sequence of
singly occupied sites ($\sigma$), holes ($\circ$), and double occupancies
($\bullet$) from site $1$ to~$L$, e.g.,
\begin{equation}
|\Psi\rangle = |\uparrow_1, \bullet_2, \circ_3,\circ_4,\downarrow_5,
\ldots\uparrow_{L-3},\bullet_{L-2},\downarrow_{L-1},\bullet_{L}\rangle
\quad .
\label{defofstate}
\end{equation}
We introduce the notations
\begin{mathletters}
\begin{eqnarray}
|\ldots,(\uparrow_l,\downarrow_{l+1}\pm \downarrow_{l},\uparrow_{l+1}),
\ldots\rangle
&=& |\ldots, \uparrow_l, \downarrow_{l+1},\ldots \rangle
\pm |\ldots,\downarrow_{l},\uparrow_{l+1},\ldots\rangle\\[3pt]
|{\rm S}_{l,l+1}=1,S^z_{l,l+1}=1\rangle&=& |\ldots,\uparrow_l,\uparrow_{l+1},
\ldots\rangle\\[3pt]
|{\rm S}_{l,l+1}=1,S^z_{l,l+1} =0\rangle &=& |\ldots,(\uparrow_l,\downarrow_{l+1}
+\downarrow_{l},\uparrow_{l+1}),\ldots\rangle\\[3pt]
|{\rm S}_{l,l+1}=1,S^z_{l,l+1}=-1\rangle &=& |
\ldots,\downarrow_l,\downarrow_{l+1},
\ldots\rangle\\[3pt]
|{\rm S}_{l,l+1}=0,S^z_{l,l+1} =0\rangle &=&
|\ldots,(\uparrow_l,\downarrow_{l+1}
-\downarrow_{l},\uparrow_{l+1}),\ldots\rangle
\end{eqnarray}
as the local spin triplet and spin singlet states.
Furthermore,
\begin{equation}
|C_{l,l+1}=1,C^z_{l,l+1}=0\rangle
= |\ldots,(\bullet_{l},\circ_{l+1} - \circ_l,\bullet_{l+1}),\ldots\rangle
\end{equation}
\end{mathletters}%
denotes the local charge
triplet state since
$\hat{C}^+|C_{l,l+1}=1,C^z_{l,l+1}=0\rangle \neq 0$.
With these definitions one finds
\begin{mathletters}
\begin{eqnarray}
\hat{\jmath}_{\rm inter,+}^{\rm H} |{\rm S}_{l,l+1}=1\rangle &=&0 \\[6pt]
\hat{\jmath}_{\rm inter,+}^{\rm H} |{\rm S}_{l,l+1}=0\rangle &=&
-itea (1+(-1)^l\delta)(1+(-1)^l\eta)(-2)
|C_{l,l+1}=1,C^z_{l,l+1}=0\rangle \; .
\end{eqnarray}
\end{mathletters}%
It is thus seen that
$\hat{\jmath}_{\rm inter}^{\rm H}$ preserves the spin of a neighboring pair
such that $\Delta S=\Delta S^z=0$ is the selection rule for the spin sector.
The selection rule for the charge sector is found as
$\Delta C=1$, $\Delta C^z=0$.
Note that the current operator does {\em not\/}
commute with~$\hat{C}^{\rm \pm}$ as defined in eq.~(\ref{defofthecs}).

Finally, the current operator is invariant against translations
by one unit cell and thus preserves the total momentum
modulo a reciprocal lattice vector ($Q=2\pi/a$ for~$\delta=0$,
$Q=\pi/a$ for~$\delta\neq 0$).
However, the current operator can
create or destroy a charge excitation with momentum~$q$,
and create or destroy a spin excitation with momentum~$-q$.
Although there is charge-spin separation in the Hubbard model
for strong coupling the current operator mixes both degrees of freedom.
This renders the calculation of the optical absorption
of the Hubbard model a very difficult problem even in the limit
of strong correlations.

\section{Exact solution of the Harris-Lange model}
\label{exactsolution}

\subsection{Translational invariant case}

The Harris-Lange model
can exactly be solved by an explicit construction of all eigenstates. 
This has recently been shown by 
(de Boer, Korepin, and Schadschneider 1995) and                    
(Schadschneider 1995) for periodic, and by (Aligia and Arrachea 1994)
for open boundary conditions.
Since optical excitations conserve total momentum we work with periodic
boundary conditions where the total momentum is a good quantum number.

The number~$N_S$ of sites with spin (singly occupied sites),
and the number~$N_C=N_d+N_h$ of sites with charge (double occupancies
and holes) are separately conserved in the Harris-Lange model.
We have $N_C+N_S=L$ lattice sites, $N=N_S+2N_d$ electrons,
and choose~$L$ to be even such that our lattice is bipartite for all~$L$.
In the sequence of singly occupied sites,
double occupancies and holes of the state~$|\Psi\rangle$
in eq.~(\ref{defofstate})
we may identify subsequences for the spins and the
charges only (independent of the position on a special site).
Additionally,  the indices $l_j$ indicate the actual position
of the charges $C_j$.
The positions occupied by the
spins are then the ones left over by the charges:
\begin{mathletters}
\begin{equation}
|\Psi\rangle=
|  ({l_1},{l_2}, \ldots{l_{N_{C}-1}},{l_{N_C}});
(C_{1},C_{2},\ldots C_{N_{C}-1},C_{N_C});
 (S_{1},S_{2},\ldots S_{N_{S}-1},S_{N_S})\rangle
\end{equation}
where in our example
\begin{equation}
(S_{1},S_{2},\ldots S_{N_{S}-1},S_{N_S})=
(\uparrow,\downarrow,\ldots\uparrow,\downarrow) 
\end{equation}
is the subsequence for the spins and
\begin{equation}
(C_{1},C_{2},C_{3},\ldots C_{N_{C}-1},C_{N_C})=
(\bullet,\circ,\circ,\ldots\bullet,\bullet) 
\end{equation}
for the charges. The sequence for the positions occupied by charges is
\begin{equation}
({l_1},{l_2},{l_3},\ldots
{l_{N_C}-1},{l_{N_C}})=(2,3,4,\ldots L-2,L) \qquad .
\end{equation}
\end{mathletters}%
Since there is nearest-neighbor hopping only
both the spin and charge
subsequences are separately {\em conserved\/}
up to cyclic permutations due to the periodic boundary conditions.

To include the boundary effect we follow 
(de Boer {\em et al.} 1995)
and (Schadschneider 1995)                                                 
and introduce the properly symmetrized spin and charge sequences.
To this end we define the operator for a cyclic permutation
of the spin sequence,
\begin{equation}
\hat{\cal T}_S (S_1,S_2,\ldots S_{N_S})
= (S_{N_S},S_1,\ldots S_{N_S-1}) \quad ,
\end{equation}
and, equivalently, $\hat{\cal T}_C$ for the charge sequence.
Let $K_S$ and $K_C$ be the smallest positive integers such that
$\left(\hat{\cal T}_S\right)^{K_S}$ and $\left(\hat{\cal T}_C\right)^{K_C}$
act as identity operators on a given spin  and charge sequence,
respectively.
Then we define the~$K_S K_C$ states
{\arraycolsep=0pt\begin{eqnarray}
&&\sqrt{K_SK_C} |({l_1}, \ldots {l_{N_C}});
(C_1,\ldots C_{N_C})_{k_C};
(S_1,\ldots S_{N_S})_{k_S}\rangle
=  \label{realspace}\\[9pt]
&& e^{\sum_l i\pi l \hat{D}_{l}}
\sum_{\nu_S=0}^{K_S-1} \sum_{\nu_C=0}^{K_C-1}
e^{ia(\nu_S k_S+\nu_C k_C)}
\hat{\cal T}_S^{\nu_S} \hat{\cal T}_C^{\nu_C}
|({l_1}, \ldots {l_{N_C}});
(C_1,\ldots C_{N_C});(S_1,\ldots S_{N_S})\rangle
\nonumber
\end{eqnarray}}%
with the momentum shifts $k_S=2\pi m_S/(K_S a)$, $m_S=0,1,\ldots (K_S-1)$,
$k_C=2\pi m_C/(K_C a)$, $m_C=0,1,\ldots (K_C-1)$.
An extra phase factor~$(-1)^{l_j}$
for each double occupancy at site~$l_j$ has been included here
through the operator
$\exp(\sum_l i\pi l \hat{D}_{l})$.
This allows to make direct contact
to the model considered by (de Boer {\em et al.} 1995);  
(Schadschneider 1995), and (Aligia {\em et al.} 1994).
There the hopping amplitudes for lower and upper Hubbard band have
opposite signs.

We transform the states of eq.~(\ref{realspace})
into momentum space with respect to the charge coordinates.
The exact eigenstates can now be classified 
according to their
spin and charge sequence, their momentum shifts $k_C$, $k_S$,
and~$N_C$ momenta from the set 
of~$k_{j}=2\pi m_j/(L a)$, $m_j=-(L/2),\ldots (L/2)-1$.
The normalized eigenstates read
{\arraycolsep=0pt\begin{eqnarray}
&&L^{N_C/2}|k_1,\ldots k_{N_C};
(C_1,\ldots C_{N_C})_{k_C};
(S_1,\ldots S_{N_S})_{k_S}\rangle
=  \label{eigenstates} \\[9pt]
&&\sum_{l_1< \ldots <l_{N_C}} \!\! \sum_{{\cal P}}(-1)^{\cal P}
\exp \! \left(ia\sum_{j=1}^{N_C} l_{{\cal P}(j)}\left(k_j+\Phi_{CS}\right) \!
\right) \!
|({l_1}, \ldots  {l_{N_C}});
(C_1,\ldots C_{N_C})_{k_C};
(S_1,\ldots S_{N_S})_{k_S}\rangle
\nonumber
\end{eqnarray}}%
where the permutations ${\cal P}$ generate a simple Slater determinant
for the momenta and the positions of the
$N_C$ charges, and $\Phi_{CS}=(k_C-k_S)/L$ is an additional momentum shift
which vanishes in the thermodynamical limit.

It is straightforward 
(de Boer {\em et al.} 1995); (Schadschneider 1995)     
but lengthy to explicitly show that
the states in eq.~(\ref{eigenstates}) have energy and momentum
\begin{mathletters}
\begin{eqnarray}
E&=&\sum_{j=1}^{N_C} \epsilon(k_j+\Phi_{CS}) +U N_d
\label{eigenenHL}\\[3pt] 
P&=& k_S +(\pi/a) (N_d-1)+ \sum_{j=1}^{N_C} (k_j+\Phi_{CS})
+ (N {\rm \ mod\ } 2) \pi/a
\qquad {\rm mod\ }2\pi/a 
\label{eigenmomHL}
\end{eqnarray}
\end{mathletters}%
with $\epsilon(k)$ given by equation~(\ref{gl3a}).
The essential arguments are repeated in the appendices~\ref{appmomentum}
and~\ref{appenerg}.

The above set of eigenstates is complete.
After summing over the subspaces with different~$K_S$,
$K_C$ the number of states which become degenerate in energy
in the thermodynamical limit is given by
$2^{N_S} N_C!/(N_d!N_h!)$. The number of possible choices for the
momenta is~$L!/(N_C!(L-N_C)!)$ since the momenta are those of a
gas of spinless Fermions. Altogether one finds for an even number of
electrons~$N$
\begin{equation}
\sum_{N_d=0}^{N/2} {N_C \choose N_d} 2^{N-2N_d} {L \choose {L-N+2N_d}}
= {2L \choose N}
\end{equation}
where eq.~(10.33.5) of (Hansen 1975) and eq.~(22.2.3) of       
(Abramovitz and Stegun 1970) have been used.                   
This exhausts the Hilbert space for fixed number of electrons~$N$.

\subsection{Dimerized Harris-Lange model}

The Harris-Lange model can also be solved in the
presence of a finite lattice distortion as long as
there is hopping between nearest neighbors only.
For the $U=\infty$ Hubbard model at less
than half filling this has already been realized
some time ago (Bernasconi, Rice, Schneider, and Str\"{a}\ss ler 1975).   
The exact eigenenergies are those of spinless Fermions
on a dimerized chain.

For the Harris-Lange model
with~$N_C$ charge excitations
we choose momenta~$|k_j|\leq \pi/(2a)$ of the reduced Brillouin zone,
and one of the~$2^{N_C}$ sequences $(\tau_1,\ldots \tau_{N_C})$
with~$\tau_j=\pm 1$.
Let us introduce the operator~$\hat{\Pi}_r$ by
{\arraycolsep=0pt\begin{eqnarray}
\hat{\Pi}_r |k_1,\ldots k_r,\ldots  k_{N_C};  &&
(C_1,\ldots C_{N_C})_{k_C};
(S_1,\ldots S_{N_S})_{k_S}\rangle
\nonumber \\[6pt]
&& = |k_1,\ldots k_r+\pi/a,\ldots k_{N_C}; 
(C_1,\ldots C_{N_C})_{k_C};
(S_1,\ldots S_{N_S})_{k_S}\rangle
\end{eqnarray}}%
and the four functions~$\xi_1^r(1)=\beta_{k_r}$,
$\xi_2^r(1)=i\alpha_{k_r}$,
$\xi_1^r(-1)=\alpha_{k_r}$, and $\xi_2^r(-1)=-i\beta_{k_r}$,
compare eq.~(\ref{alphabeta}).
The eigenstates for fixed number of charge excitations~$N_C$
can then be written as
\begin{eqnarray}
|k_1,\tau_1;\ldots  &&k_{N_C},\tau_{N_C};
(C_1,\ldots C_{N_C})_{k_C};
(S_1,\ldots S_{N_S})_{k_S}\rangle
= \nonumber \\[6pt]
&&\left[
\prod_{r=1}^{N_C} \left( \xi_1^r(\tau_r) + \xi_2^r(\tau_r)\hat{\Pi}_r\right)
\right] 
|k_1,\ldots  k_{N_C}; 
(C_1,\ldots C_{N_C})_{k_C};
(S_1,\ldots S_{N_S})_{k_S}\rangle \; .
\end{eqnarray}
This state corresponds to $(N_C+\sum_r\tau_r)$ ($(N_C-\sum_r\tau_r)/2)$
spinless Fermions in the upper (lower) Peierls subband.

The corresponding energies and momenta of these states
are obviously given by
\begin{mathletters}
\begin{eqnarray}
E&=&\sum_{j=1}^{N_C} E(k_j+\Phi_{CS})\tau_j +U N_d\\[3pt] 
P&=& k_S + \sum_{j=1}^{N_C} (k_j+\Phi_{CS})
\quad {\rm mod\ }\pi/a 
\end{eqnarray}
\end{mathletters}%
where~$E(k)$ has been given in eq.~(\ref{peierlsen}).

\subsection{Band picture interpretation of the spectrum}
\label{bandpictureinterpretation}

\subsubsection{Translational invariant case}

The exact solution for the Harris-Lange model
can be interpreted in terms of upper and lower Hubbard bands.
To simplify the discussion we will ignore the momentum shift~$\Phi_{CS}$
in this subsection.
For $U>W$ the ground state of the half-filled band~$N=L$ has energy zero
and is $2^L$-fold degenerate, and we may choose the fully polarized
ferromagnetic state as our reference state,
$|{\rm FM}\rangle=|\uparrow,\ldots  \uparrow\rangle$.
Note that this state has momentum~$\pi/a$
on a chain with an even number of sites~$L$, see eq.~(\ref{eigenmomHL}).

We may now add an electron.
We obtain all exact eigenstates for $N=L+1$ electrons, $N_d=1$, and
all spins up as
\begin{mathletters}
\begin{equation}
|k;(\bullet)_{k_C=0},(\uparrow,\ldots  \uparrow)_{k_S=0}\rangle
= \hat{c}_{k,\downarrow}^+ |{\rm FM}\rangle
\label{plusone}
\end{equation}
with momentum~$P=k+\pi/a$ and energy~$E=\epsilon(k)+U$.
The state in eq.~(\ref{plusone}) is interpreted
as a {\em particle\/} at momentum~$k$ in the upper Hubbard
band which has the dispersion relation~$\epsilon(k)+U$.
The momentum~$\pi/a$ is attributed to the ferromagnetic reference state.

We may equally well take out an electron from the fully polarized state.
We obtain all exact eigenstates for $N=L-1$ electrons, $N_h=1$,
and all spins up as
\begin{equation}
|k;(\circ)_{k_C=0},(\uparrow,\ldots  \uparrow)_{k_S=0}\rangle
= - \hat{c}_{\pi/a-k,\uparrow} |{\rm FM}\rangle
\label{minusone}
\end{equation}
\end{mathletters}%
with momentum~$P=-(\pi/a-k)+\pi/a$ and energy~$E=\epsilon(k)$.
Note that the states of eq.~(\ref{minusone})
and those of eq.~(\ref{plusone})
can be generated from each other by the
particle-hole transformation of eq.~(\ref{parthole}).
Their momenta differ by~$\pi/a$
since their respective numbers of double occupancies differ by one.

Since the ground state corresponds to
a completely filled lower Hubbard band
we {\em interpret\/} the state in eq.~(\ref{minusone})
as a hole in the lower Hubbard band at~$k_h=\pi/a-k$,
and the momentum~$\pi/a$ is again attributed to the ferromagnetic reference
state. 
The lower Hubbard band must have the dispersion relation~$\epsilon(k)$
for {\em particles\/} because a {\em hole\/} at $k_h=\pi/a-k$
has momentum~$P=-(\pi/a-k)$ and energy~$E=-\epsilon(k_h)=
-\epsilon(-k+\pi/a)=\epsilon(k)$.

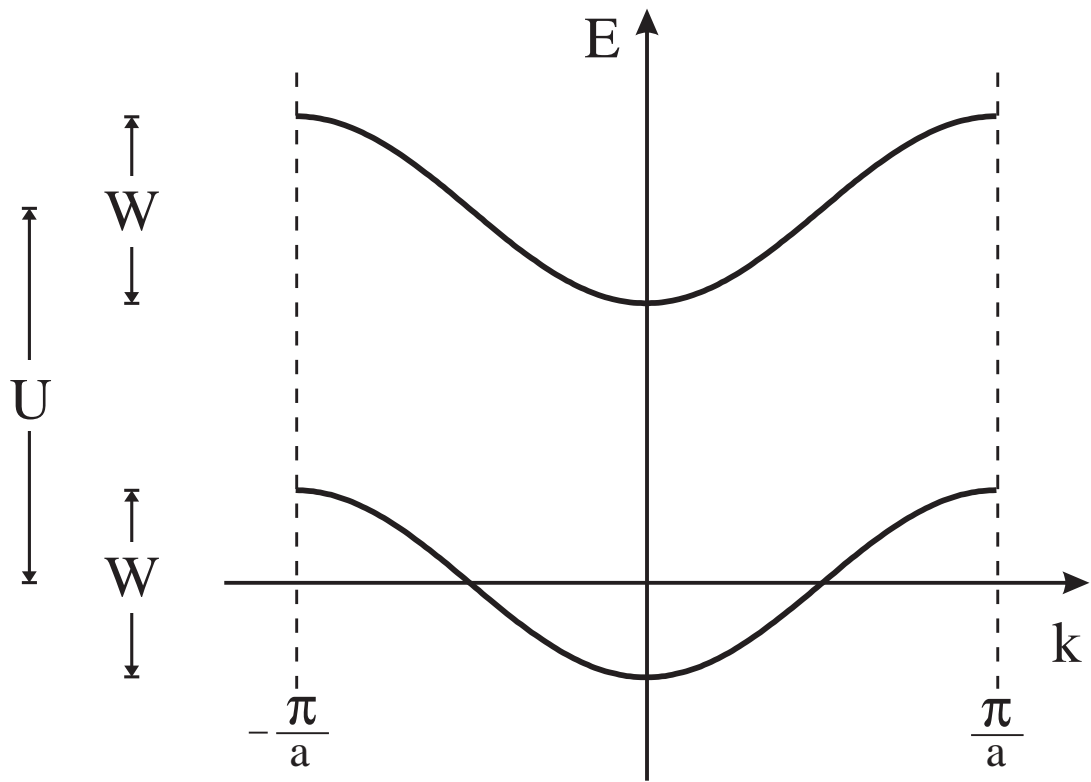
\begin{figure}[th]
\caption{Band structure interpretation of the exact eigenenergies
of the Harris-Lange model for U=2W.}
\label{HarrisLangedis}
\end{figure}

The band structure for the Harris-Lange model is
depicted in figure~\ref{HarrisLangedis}.
It displays the parallel upper and lower Hubbard bands
with band width~$W$ separated by a distance~$U$.
It is amusing that the celebrated Hubbard-I
approximation (Hubbard 1963); (Mazumdar and Soos 1981)             
also gives parallel bands. Those bands, however, carry a spin index
while charge-spin separation is most essential in one dimension.
Furthermore, the width of those bands is only {\em half\/}
of the exact band width~$W$.

Our band structure picture has to be used carefully if there are
more than one double occupancy or hole. Figure~\ref{HarrisLangedis} suggests
that there are~$L$ states available {\em both\/} in the upper {\em and\/}
in the lower Hubbard band, altogether~$2L$ independent states.
However, this cannot be the case because for~$N_d=N_h=L/2$
we would have~${L \choose L/2}{L \choose L/2}$ states in the band picture
while the Hilbert space actually has only the dimension~${L \choose L/2}$.
The exact solution shows how an appropriate exclusion
principle between particles in the upper Hubbard band
and holes in the lower Hubbard band can be formulated.

For fixed spin background and fixed $k_1$, $k_2$
there are four exact eigenstates with two charges at~$k_1\neq k_2$.
They all have the kinetic energy~$T=\epsilon(k_1)+\epsilon(k_2)$.
They correspond to four different charge excitations in the band picture:
(i)~two particles at momenta~$k_1$, $k_2$ in the upper Hubbard band,
(ii)~two holes at momenta~$\pi/a-k_1$, $\pi/a-k_2$ in the
lower Hubbard band, (iii)~a particle at momentum~$k_1$
in the upper Hubbard band
and a hole at momentum~$\pi/a-k_2$ in the lower Hubbard band,
(iv)~a particle at momentum~$k_2$ in the upper Hubbard band
and a hole at momentum~$\pi/a-k_1$ in the lower Hubbard band.
The condition~$k_1\neq k_2$ is naturally
fulfilled in cases~(i) and~(ii), if we assign a fermionic character to
our particles in the upper and holes in the lower Hubbard band, respectively.
In case~(iii), however, we have to explicitly {\em demand\/}
that the momentum at which we create the hole, $k_h=\pi/a-k_2$,
fulfills~$k_1\neq k_2$, i.~e., $k_h\neq \pi/a-k_1$.
This is the same condition which results from case~(iv).

We thus see that a particle in the upper Hubbard band at momentum~$k$
actually blocks the creation of a hole in the lower Hubbard band at
momentum~$\pi/a-k$ (this is probably the simplest example of a ``statistical
interaction'', see (Haldane 1991)).                
With this additional rule the counting of states
in the band picture is correct, and the band picture interpretation
gives indeed the {\em exact\/} results for the Harris-Lange model.
The effective Hamiltonian for fermionic particles in the upper
($\hat{u}_k$) and lower ($\hat{l}_k$) Hubbard band thus reads
\begin{mathletters}
\label{effHL}
\begin{eqnarray}
\hat{H}_{\rm HL}^{\rm band} &=& \hat{P}_{ul}  \sum_{|k|\leq \pi/a}\left[
(U+\epsilon(k)) \hat{n}_{k}^{u} + \epsilon(k) \hat{n}_{k}^{l} \right]
\hat{P}_{ul}
\\[6pt]
\hat{P}_{ul} &=& \prod_{|k|\leq \pi/a}
\left[ 1 -\left(1- \hat{n}_{\pi/a-k}^{l}\right)
\hat{n}_{k}^{u} \right]
\end{eqnarray}
\end{mathletters}%
with $\hat{n}_{k}^{u}=\hat{u}_{k}^+  \hat{u}_{k}^{\phantom{+}}$,
$\hat{n}_{k}^{l}=\hat{l}_{k}^+  \hat{l}_{k}^{\phantom{+}}$.
The projection operators guarantee that there is no hole in the
lower Hubbard band at momentum~$\pi/a-k$,
if there is already a particle at momentum~$k$ in the upper Hubbard band.
For half-filling at zero temperature the lower Hubbard band is completely
filled.

\subsubsection{Dimerized Harris-Lange model}

The case of the dimerized Hubbard model can be treated accordingly.
The upper and lower Hubbard band now split into two Peierls subbands
with dispersion relations~$\pm E(k)$.
Formally the band structure Hamiltonian for the lower band becomes
(compare eq.~(\ref{Tink}))
\begin{equation}
\hat{T}_{\rm LHB}^{\rm band}(\delta)= \sum_{|k|\leq \pi/(2a)}
\epsilon(k)\left(
\hat{l}_{k}^+\hat{l}_{k}^{\phantom{+}}
-
\hat{l}_{k+\pi/a}^+\hat{l}_{k+\pi/a}^{\phantom{+}}
\right)
-i \Delta(k)
\left( \hat{l}_{k+\pi/a}^+\hat{l}_{k}^{\phantom{+}}
-
\hat{l}_{k}^+\hat{l}_{k+\pi/a}^{\phantom{+}}
\right) \; ,
\label{Tdimnotdiag}
\end{equation}
and a similar expression holds for the upper Hubbard band.

The band picture Hamiltonian can easily be brought into diagonal
form as in the Peierls case. The quasi-particles in the four subbands
are finally described by
\begin{eqnarray}
\hat{H}_{\rm HL}^{\rm dim,\, band} &=& \hat{P}_{u^+l^+}\hat{P}_{u^-l^-}
 \sum_{|k|\leq \pi/(2a)}\biggl[
(U+E(k)) \hat{n}_{k,+}^{u} + (U-E(k)) \hat{n}_{k,-}^{u}
\nonumber \\[6pt]
&& \phantom{\hat{P}_{u^+l^+}\hat{P}_{u^-l^-}
 \sum_{|k|\leq\pi/(2a)}\biggl[}
+ E(k) \hat{n}_{k,+}^{l} -E(k) \hat{n}_{k,-}^{l}
\biggr] \hat{P}_{u^+l^+}\hat{P}_{u^-l^-}
\label{bandhldim}
\\[6pt]
\hat{P}_{u^{\pm}l^{\pm}} &=& \prod_{|k|\leq\pi/(2a)} \left[ 1- \left(1-
\hat{n}_{-k,\pm}^{l}\right)\hat{n}_{k,\pm}^{u}\right] \nonumber 
\end{eqnarray}
with $\hat{n}_{k,\pm}^{u}=\hat{u}_{k,\pm}^+  \hat{u}_{k,\pm}^{\phantom{+}}$,
$\hat{n}_{k,\pm}^{l}=\hat{l}_{k,\pm}^+  \hat{l}_{k,\pm}^{\phantom{+}}$
as the number operators for
the quasi-particles for the upper ($\tau=+$) and lower ($\tau=-$)
Peierls subband in the upper~($u$) and lower~($l$) Hubbard band.
The quasi-particles in each subband obey a fermionic
exclusion principle in the same Hubbard band. 
In addition, a particle at momentum~$k$
in the upper Hubbard band in the upper (lower) Peierls subband blocks the
creation of a hole at momentum~$-k$ in the lower Hubbard band in
the upper (lower) Peierls subband.
There is no hole in the lower Hubbard band at momentum~$-k$
in the upper or lower Peierls subband,
if there is already a particle at momentum~$k$ in the upper Hubbard band
in the corresponding Peierls subband.
Note that the reciprocal lattice vector is
now given by~$\pi/a$.
Thereby the projection operators guarantee the proper counting
of states.

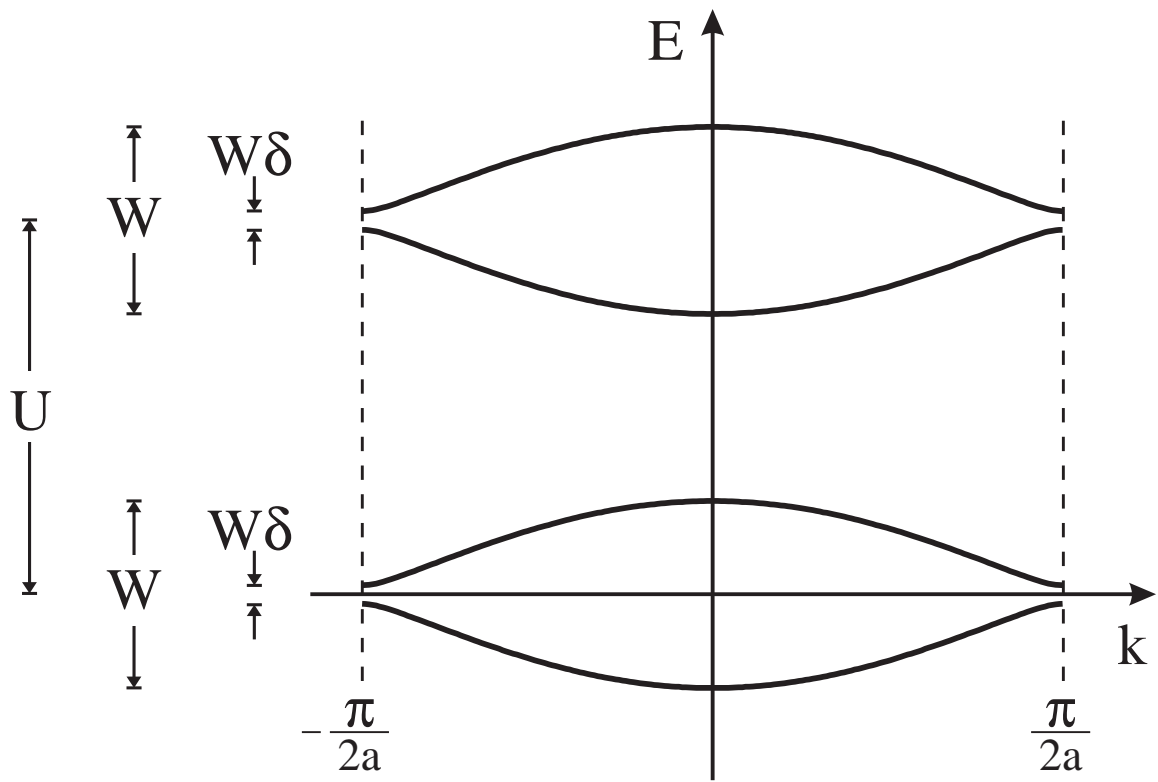
\begin{figure}[th]
\caption{Band structure interpretation of the exact eigenenergies
of the dimerized Harris-Lange model for U=2W, $\delta=0.2$.}
\label{HueckelHarrisLangedis}
\end{figure}

The resulting band structure is shown in figure~\ref{HueckelHarrisLangedis}.
The upper and lower Hubbard band are both Peierls split and display
the Peierls gap~$W\delta$ at the zone boundaries~$\pm \pi/(2a)$.
Note that the upper (lower) Peierls subbands
are still parallel.

\subsection{Band picture interpretation of the current operator}

\subsubsection{Translational invariant case}

According to the spectral decomposition of the current-current correlation
function, eq.~(\ref{decomp}), we need to determine the excitation
energy~$E_n-E_0$ of an exact eigenstate~$|n\rangle$
and its oscillator strength~$|\langle 0 | \hat{\jmath}|n\rangle|^2$.
The respective total momenta of these states are~$P_0$ and~$P_n$.

We are interested in optical excitations from a state with 
singly occupied sites only. The excited states which can be reached from
this state have one pair of hole and double occupancy, i.~e.,
\begin{mathletters}
\begin{eqnarray}
|0\rangle &=& |(S_1,\ldots S_L)_{k_S}\rangle\\[3pt]
|n\rangle &=& |k_1,k_2; (\bullet \circ)_{k_C=0};
(S_1,\ldots S_{L-2})_{k_S^{\prime}}\rangle
\end{eqnarray}
\end{mathletters}%
where we used the fact that~$\hat{\jmath}$ creates a charge triplet
with~$k_C=0$. Note that~$(k_1,k_2)$ is the same state as~$(k_2,k_1)$.
We denote~$k_1=k+q/2$, $k_2=\pi/a-k+q/2$ since we will finally represent
the state~$|n\rangle$
by a particle in the upper Hubbard band at momentum~$k+q/2$ and a hole
in the lower Hubbard band at momentum~$k-q/2$. 

Since~$\hat{\jmath}$ conserves the total momentum we already know
that~$P_0=P_n$ which implies
$k_S=q+k_S^{\prime}(L-2)/L$, see eq.~(\ref{eigenmomHL}).
Hence, $k_S^{\prime}=L(k_S-q)/(L-2)$ has to hold. Recall that~$k_S^{\prime}$
is quantized in units of~$2\pi/((L-2)a)$.
These considerations imply that the charge (spin) sector in~$|n\rangle$
carries momentum~$q$ ($-q$) relative to~$|0\rangle$.
In the thermodynamical limit the excitation energy is given by
\begin{equation}
E(k,q)=U + \epsilon(k+q/2) - \epsilon(k-q/2)= U + 4t\sin(ka)\sin(qa/2)\; .
\label{EKQ}
\end{equation}
Note that the excitation energy does not depend on the spin configuration.
For this reason it is possible to find a formally equivalent band picture
for the charge sector alone
that gives the same optical absorption as the original model.
Since~$\hat{\jmath}$ itself carries all the information on the 
conservation laws
(momentum, charge, and spin quantum numbers) we may equally well
work with the (normalized) states
\begin{eqnarray}
|k+\frac{q}{2};\frac{\pi}{a}-k+\frac{q}{2}\rangle &=& 
\frac{1}{L} \sum_{l_1<l_2}
\left( e^{i(k+q/2)l_1a}e^{i(\pi/a-k+q/2)l_2a}
 -e^{i(k+q/2)l_2a}e^{i(\pi/a-k+q/2)l_1a} \right)
 \nonumber \\
  && \phantom{\frac{1}{L} \sum_{l_1<l_2}}
  (-1)^{l_1} |S_1^{\prime},\ldots S_{l_1-1}^{\prime},\bullet_{l_1},
S_{l_1}^{\prime},\ldots S_{l_2-2}^{\prime}, \circ_{l_2},
S_{l_2-1}^{\prime},\ldots S_{L-2}^{\prime} \rangle 
\end{eqnarray}
rather than the exact eigenstates of eq.~(\ref{eigenstates}).
This will simplify the notation since we do not have to take
any summation restrictions into account.

For fixed~$(k,q)$ and fixed spin configuration~$(S_1^{\prime},\ldots
S_{L-2}^{\prime})$ we calculate
\begin{eqnarray}
\langle 0 |\hat{\jmath}_{\rm inter,-}^{\rm H}
|k+\frac{q}{2};\frac{\pi}{a}-k+\frac{q}{2}\rangle &=&
-iea\epsilon(k) e^{iqa/2}  \\
&& \frac{1}{L} \sum_l e^{iqla} 
\langle 0 |S_1^{\prime},\ldots S_{l-1}^{\prime},
\left( \uparrow_{l}\downarrow_{l+1}-\downarrow_{l}\uparrow_{l+1}\right),
S_{l}^{\prime},\ldots S_{L-2}^{\prime} \rangle \; .
\nonumber
\end{eqnarray}
We define the operators~$\hat{x}_q^+$ and $\hat{x}_q^{\phantom{+}}$ via their
product
\begin{eqnarray}
\hat{x}_q^+ \hat{x}_q^{\phantom{+}} &=&
\sum_{S_1^{\prime},\ldots S_{L-2}^{\prime}}
\frac{1}{L^2} \sum_{l,r} e^{iq(l-r)a}
\langle 0 | S_1^{\prime},\ldots S_{l-1}^{\prime},
\left( \uparrow_{l}\downarrow_{l+1}-\downarrow_{l}\uparrow_{l+1}\right),
S_{l}^{\prime},\ldots S_{L-2}^{\prime} \rangle \label{xqxq}\\[6pt]
&&
\phantom{\sum_{S_1^{\prime},\ldots S_{L-2}^{\prime}}
\frac{1}{L^2} \sum_{l,r} e^{iq(l-r)} }
\langle S_{L-2}^{\prime},\ldots S_{r}^{\prime},
\left( \downarrow_{r+1} \uparrow_{r}-\uparrow_{r+1}\downarrow_{r}\right),
S_{r-1}^{\prime},\ldots S_{1}^{\prime} | 0 \rangle \; . \nonumber
\end{eqnarray}
Summed over all intermediate spin configurations
the oscillator strength for fixed $(k,q)$ now becomes
\begin{equation}
\Bigl|\langle 0 |\hat{\jmath}_{\rm inter,-}^{\rm H}
|k+\frac{q}{2};\frac{\pi}{a}-k+\frac{q}{2}\rangle\Bigr|^2 =
\Bigl|-iea \epsilon(k)\Bigr|^2 \hat{x}_q^+\hat{x}_q^{\phantom{+}} \; .
\end{equation}
It is clear that we have hidden a very difficult many-body problem in
the operators~$\hat{x}_q$.

Nevertheless we are now in the position to identify the interband
current operator in the band picture.
It is given by
\begin{equation}
\hat{\jmath}_{\rm inter}^{\rm band}=
\sum_{|k|,|q|\leq \pi/a}
iea\epsilon(k) \left( \hat{u}_{k+q/2}^+\hat{l}_{k-q/2}^{\phantom{+}}
\hat{x}_{q}^{\phantom{+}} - \hat{l}_{k-q/2}^+\hat{u}_{k+q/2}^{\phantom{+}}
\hat{x}_{q}^+ \right)\; .
\label{jhlbandpicture}
\end{equation}
This operator acts in the same space as the band Hamiltonian of
section~\ref{bandpictureinterpretation}.
It is seen that the condition~$k+q/2 \neq \pi/a-(k-q/2)$ is automatically
fulfilled since~$\epsilon(\pi/(2a))=0$. Consequently, 
the projection operators
in eq.~(\ref{effHL}) can again
be ignored for the case of linear optical absorption.

\subsubsection{Dimerized Harris-Lange model}

For a distorted lattice
the current operator can also modify the momentum of the state
by $\pi/a$.
Thus we address {\em four\/} possible states for fixed~$(k,q)$ from 
the reduced Brillouin zone,
$|k+q/2;\pi/a-k+q/2\rangle$,
$|k+q/2;-k+q/2\rangle$,
$|\pi/a+k+q/2;\pi/a-k+q/2\rangle$, and 
\hbox{$|\pi/a+k+q/2;-k+q/2\rangle$}.
The same analysis as in the previous subsection leads us to the
definition of the operators~$\hat{x}_q^{+}(\delta,\eta)$,
$\hat{x}_q^{\phantom{+}}(\delta,\eta)$ with the property
\newpage
\typeout{forced newpage}
{\arraycolsep=0pt\begin{eqnarray}
\hat{x}_q^{+}(\delta,\eta)\hat{x}_{q'}^{\phantom{+}}(\delta,\eta)
&=& 
\sum_{S_1^{\prime},\ldots S_{L-2}^{\prime}}
\frac{1}{L^2} \sum_{l,r} e^{i(ql-q'r)a}
\bigl(1+\eta\delta +(-1)^l(\delta+\eta)\bigr)
\bigl(1+\eta\delta +(-1)^r(\delta+\eta)\bigr)
\nonumber \\[6pt]
&& \phantom{\sum_{S_1^{\prime},\ldots S_{L-2}^{\prime}}
\frac{1}{L^2} \sum_{l,r} e^{i(ql-q'r)} }
\langle 0 | S_1^{\prime},\ldots S_{l-1}^{\prime},
\left( \uparrow_{l}\downarrow_{l+1}-\downarrow_{l}\uparrow_{l+1}\right),
S_{l}^{\prime},\ldots S_{L-2}^{\prime} \label{xqxqprime} \rangle
\\[6pt]
&& \phantom{\sum_{S_1^{\prime},\ldots S_{L-2}^{\prime}}
\frac{1}{L^2} \sum_{l,r} e^{i(ql-q'r)} }
\langle S_{L-2}^{\prime},\ldots S_{r}^{\prime},
\left( \downarrow_{r+1} \uparrow_{r}-\uparrow_{r+1}\downarrow_{r}\right),
S_{r-1}^{\prime},\ldots S_{1}^{\prime} | 0 \rangle \; .
\nonumber
\end{eqnarray}}%
In practice, $q'=q$ or $q'=q+\pi/a$.

The interband current operator becomes
\begin{eqnarray}
\hat{\jmath}_{\rm inter}^{\rm band}&=&
\hat{\jmath}_{\rm inter,+}^{\rm band}+
\hat{\jmath}_{\rm inter,-}^{\rm band} \nonumber
\\[6pt]
\hat{\jmath}_{\rm inter,+}^{\rm band}&=&
\left(\hat{\jmath}_{\rm inter,-}^{\rm band}\right)^+
\nonumber \\[6pt]
\hat{\jmath}_{\rm inter,-}^{\rm band}&=&
\sum_{|q|,|k|\leq \pi/(2a)}\biggl\{ -iea \epsilon(k) \Bigl[
\hat{l}_{k-q/2}^+ \hat{u}_{k+q/2}^{\phantom{+}} 
- \hat{l}_{k-q/2+\pi/a}^+ \hat{u}_{k+q/2+\pi/a}^{\phantom{+}}
\Bigr] \hat{x}_{q}^+ \label{jinterband}\\[9pt]
&& \phantom{\sum_{|q|,|k|\leq \pi/(2a)}\biggl\{ }
+ea \frac{\Delta(k)}{\delta} \Bigl[
\hat{l}_{k-q/2}^+ \hat{u}_{k+q/2+\pi/a}^{\phantom{+}} 
- \hat{l}_{k-q/2+\pi/a}^+ \hat{u}_{k+q/2}^{\phantom{+}}
\Bigr] \hat{x}_{q+\pi/a}^+  \biggr\} \nonumber \; .
\end{eqnarray}
Again the conditions~$k+q/2 \neq \pi/a-(k-q/2)$ and
$k+q/2 \neq -(k-q/2)$ will not be violated
since~$\epsilon(\pi/(2a))=0$ and
$\Delta(0)=0$, respectively.
Consequently, the projection operators
in eq.~(\ref{bandhldim})
can be ignored for the case of linear optical absorption.

In semiconductor physics one prefers to work with
the dipole operator rather than the current operator
to set up the perturbation theory in the electrical field
(Haug {\em et al.} 1990).              
The corresponding expressions for the dipole operator
for Hubbard interband transitions  are derived in appendix~\ref{appdipoleHL}.

\section{Optical absorption in the Harris-Lange model}
\label{optabsHL}
\subsection{Spin average}

All states with no double occupancy are possible ground states in the
Harris-Lange model at half-filling.
Instead of looking at the optical absorption for a specific state~$|0\rangle$
it is more reasonable to calculate the {\em average\/}
absorption, i.~e.,
\begin{equation}
\overline{{\rm Im}\{\chi(\omega)\}} =
\frac{1}{2^L} \sum_{|0\rangle} {\rm Im}\{\chi_{|0\rangle}(\omega)\} \; .
\end{equation}
For the Hubbard model this corresponds to 
temperatures $k_{\rm B}T \gg J={\cal O}(W^2/U)$ (``hot-spin case'').
The calculation is performed in appendix~\ref{spiav}.
We find the result
\begin{mathletters}
\begin{eqnarray}
\langle \hat{x}_q^+\hat{x}_{q'}\rangle &=&
\frac{1}{2L} \biggl\{ 
\delta_{q,q'} 
\left[(1+\delta\eta)^2 g(q)+(\delta+\eta)^2 g(q+\frac{\pi}{a})\right]
\nonumber \\[3pt]
&& \phantom{\frac{1}{2L} \biggl\{ }
+ \delta_{q,q'+\frac{\pi}{a}} (1+\delta\eta)(\delta+\eta) 
\left[g(q)+ g(q+\frac{\pi}{a})\right]
\biggr\}\label{averagexq}\\[3pt]
g(q) &=& \frac{3}{5+4\cos(qa)} \; .
\end{eqnarray}\end{mathletters}%
The spin problem could thus be traced out
completely. 
It is seen that $\hat{x}_q$ keeps its operator character
until we have expressed the current operator in terms of
the Fermion operators for the four Peierls subbands.

\subsection{Optical absorption}

\subsubsection{Translational invariant case}

The real part of the average optical conductivity becomes
\begin{equation}
{\rm Re}\{\overline{\sigma(\omega >0)} \}
= \frac{\pi {\cal N}_{\perp}}{2 L^2 a \omega} \sum_{|q|,|k|\leq \pi/a}
(ea \epsilon(k))^2 g(q) \delta(\omega - E(k,q))
\label{elliptic}
\end{equation}
with $E(k,q)$ from equation~(\ref{EKQ}). 
The above expression can be written as
\begin{equation}
\overline{\sigma_{\rm red}(\omega >0)} =
\frac{1}{4\pi}\int_{|u|}^1\frac{dx}{x^2} \frac{\sqrt{x^2-u^2}}{\sqrt{1-x^2}}
\frac{3}{9-8x^2} \label{what}
\end{equation}
for the reduced optical conductivity with $u=|\omega-U|/W\leq 1$.
This integral can be expressed as a sum over elliptic integrals 
but we rather prefer to discuss some special cases.

The optical absorption is restricted to $|\omega-U|\leq W$. 
Near the band edges the absorption increases linearly 
which can be seen from equation~(\ref{what}) by a transformation~$x\to
1-y/|u|$. 
The more interesting case is $\omega=U$. 
Now the integrand displays a $1/x$ singularity for~$|u|\to 0$.
The {\em parallel\/} Hubbard bands give rise to a logarithmic divergence, 
$\sigma(\omega\to U) \sim |\ln(\omega-U)|$, since their large
joint density of states for $\omega=U$ survives even in the presence
of a spinon bath that provides any momentum to the charge sector.

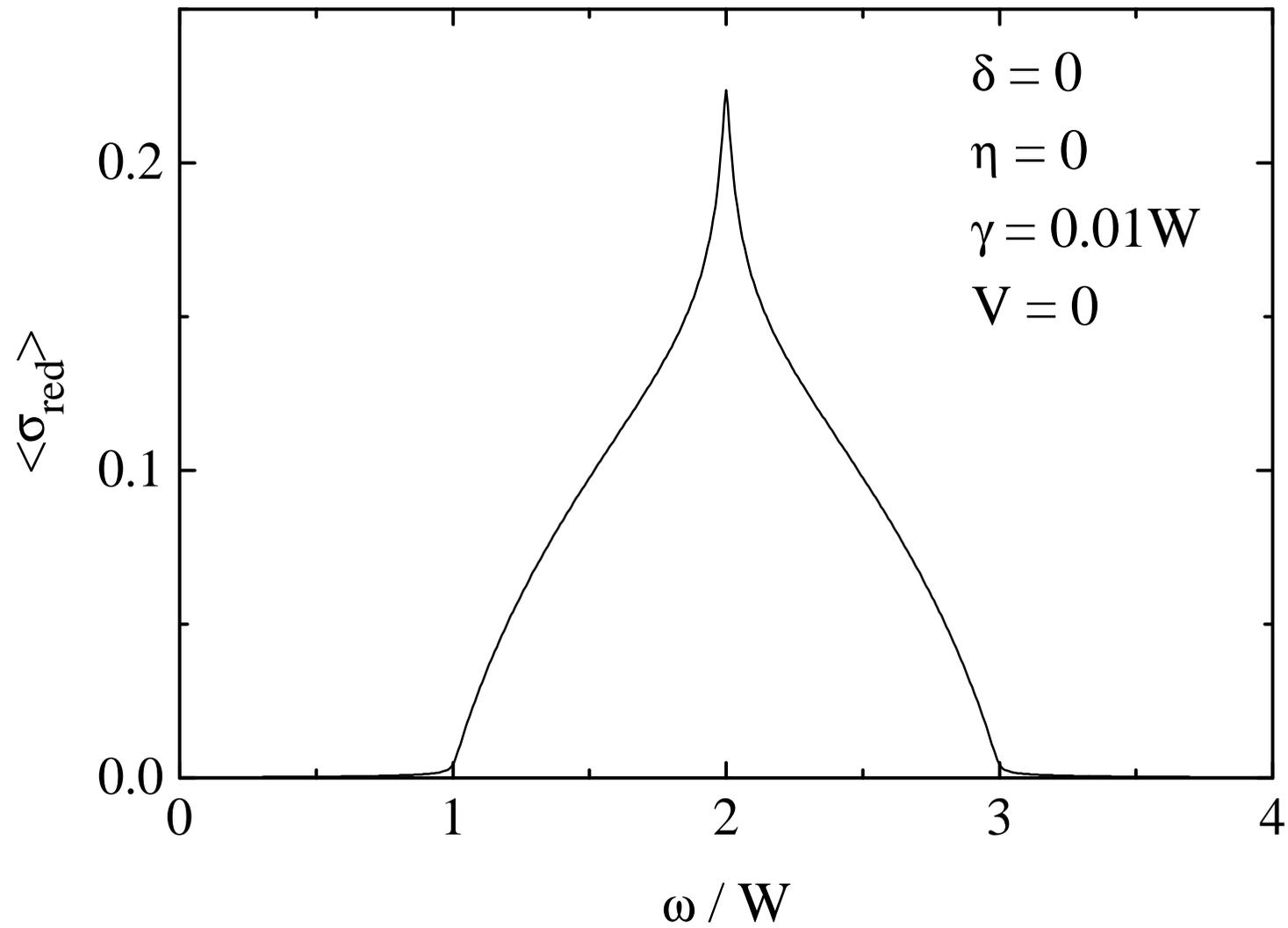
\begin{figure}[th]
\caption{Reduced average optical conductivity,
$\overline{\sigma_{\rm red}(\omega >0)}$,
in the Harris-Lange model for $U=2W$.
A broadening of~$\gamma=0.01W$ has been included.}
\label{hl00}
\end{figure}

The overall behavior of the optical absorption
is shown in figure~\ref{hl00}.
The same absorption curve has been obtained earlier in
(Lyo {\em et al.} 1977) for their ``random'' spin background.   
The result for their ``ferromagnetic'' spin background
follows when we put~$g(q)\equiv 1$, as expected.
We will comment on their N\'{e}el state results in 
(Gebhard {\em et al.} III 1996).      

\subsubsection{Dimerized Harris-Lange model}

We have to diagonalize the interband current operator of eq.~(\ref{jinterband})
in terms of the Peierls operators for the lower Hubbard band
\begin{mathletters}
\label{trafoforl}
\begin{eqnarray}
\hat{l}_{k}&=&\alpha_{k}\hat{l}_{k,-}+\beta_k\hat{l}_{k,+}\\[3pt]
\hat{l}_{k+\pi/a}&=&-i\beta_{k}\hat{l}_{k,-}+i\alpha_k\hat{l}_{k,+}
\end{eqnarray}
\end{mathletters}%
for $|k|\leq \pi/(2a)$.
The transformation for the upper Hubbard band is analogous.
With this definition the Hamiltonian in the band picture interpretation
became diagonal, see eq.~(\ref{bandhldim}).

The interband current operator becomes
\begin{equation}
\hat{\jmath}_{\rm inter,-}^{\rm band} =\sum_{\tau,\tau'=\pm 1}
\sum_{|k|,|q| \leq \pi/(2a)}
\lambda_{\tau,\tau'} (k,q)
\hat{l}_{k-q/2,\tau}^+\hat{u}_{k+q/2,\tau'}^{\phantom{+}}
\label{jinterdimHL}
\end{equation}
with
\begin{mathletters}
\label{thelambdas}
\begin{eqnarray}
\lambda_{+,+}(k,q) &=& iea \left[ \epsilon(k)
(\alpha_{+}^{\phantom{*}}\alpha_{-}^*
-\beta_{+}^{\phantom{*}}\beta_{-}^*) \hat{x}_q^+
+\frac{\Delta(k)}{\delta}
(\alpha_{+}^{\phantom{*}}\beta_{-}^*
+\beta_{+}^{\phantom{*}}\alpha_{-}^*) 
\hat{x}_{q+\pi/a}^+ \right] \\[6pt]
\lambda_{+,-}(k,q) &=& iea \left[ - \epsilon(k)
(\alpha_{+}^{\phantom{*}}\beta_{-}^*
+\beta_{+}^{\phantom{*}}\alpha_{-}^*) \hat{x}_q^+
+\frac{\Delta(k)}{\delta}
(\alpha_{+}^{\phantom{*}}\alpha_{-}^*
-\beta_{+}^{\phantom{*}}\beta_{-}^*)
\hat{x}_{q+\pi/a}^+  \right]
\end{eqnarray}
\end{mathletters}%
and $\lambda_{-,-}(k,q)= -\lambda_{+,+}(k,q)$,
$\lambda_{-,+}(k,q)=\lambda_{+,-}(k,q)$.
Here we used the short-hand notation
$\alpha_{\pm}=\alpha_{k\pm q/2}$ etc. Note that these quantities can be
complex for~$q \neq 0$.

The average optical conductivity becomes
\begin{mathletters}
\label{monstersigma}
\begin{equation}
{\rm Re}\{\overline{\sigma(\omega >0, \delta,\eta)} \}
= \frac{\pi {\cal N}_{\perp}}{L a \omega}
\sum_{\tau,\tau'=\pm 1} \sum_{|q|,|k| \leq \pi/(2a)}
\left| \lambda_{\tau,\tau'} (k,q)\right|^2
\delta(\omega - E_{\tau,\tau'}(k,q))
\end{equation}
with the absorption energies between the respective Peierls subbands
\begin{equation}
E_{\tau,\tau'}(k,q) = U +\tau' E(k+q/2)-\tau E(k-q/2) \; ,
\label{Etautauprime}
\end{equation}
see eq.~(\ref{bandhldim}) and figure~\ref{HueckelHarrisLangedis}.
The transition matrix elements are given by
\begin{eqnarray}
\left| \lambda_{\tau,\tau'} (k,q)\right|^2 &=&
 \frac{(ea)^2}{2L} \Biggl\{ 
g(q) \left| (1+\delta\eta) \epsilon(k)f_{\tau,\tau'}
+\tau\tau' (\delta+\eta) \frac{\Delta(k)}{\delta}f_{\tau,-\tau'}\right|^2
\nonumber \\[3pt]
&& \phantom{ \frac{(ea)^2}{2L} \Biggl\{  }
+ g(q+\frac{\pi}{a}) \left| (\delta+\eta) \epsilon(k)f_{\tau,\tau'}
+\tau\tau' (1+\delta\eta) \frac{\Delta(k)}{\delta}f_{\tau,-\tau'}\right|^2
\Biggr\}
\label{thelambdassquared}
\end{eqnarray}
\end{mathletters}%
with the help functions
\begin{mathletters}
\label{thefs}
\begin{eqnarray}
f_{+,+}(k,q) = f_{-,-}(k,q) &=&
\alpha_{k+q/2}^{\phantom{*}}\alpha_{k-q/2}^*
- \beta_{k+q/2}^{\phantom{*}}\beta_{k-q/2}^*
\\[3pt]
f_{+,-}(k,q) = f_{-,+}(k,q) &=&
\alpha_{k+q/2}^{\phantom{*}}\beta_{k-q/2}^*
+ \beta_{k+q/2}^{\phantom{*}}\alpha_{k-q/2}^*
\end{eqnarray}
\end{mathletters}%
where $\alpha_k$, $\beta_k$ are given in eq.~(\ref{alphabeta}).
It can easily be checked that the case~$\delta=\eta=0$ is reproduced.
For $\delta=1$, $\eta=0$ one recovers the result for
the average optical conductivity of $L/2$ independent
two-site systems since $E(k)=2t$,
$\lambda_{+,-}(k,q)=0$, and $|\lambda_{+,+}(k,q)|^2
= (2tea)^2(g(q)+g(q+\pi/a))/(2L)$:
\begin{equation}
{\rm Re}\{\overline{\sigma(\omega >0, \delta=1,\eta=0)} \}
= \frac{L}{2}  \frac{{\cal N}_{\perp}}{La\omega} \pi \delta(\omega -U)
\frac{(Wea)^2}{4}
\end{equation}
where we used $\int_{-\pi}^{\pi}dq/(2\pi)\, g(qa)=1$.
For the direct calculation we have to recall that 
only the singlet of the four spin states contributes,
and the hopping between the two sites is~$2t$.

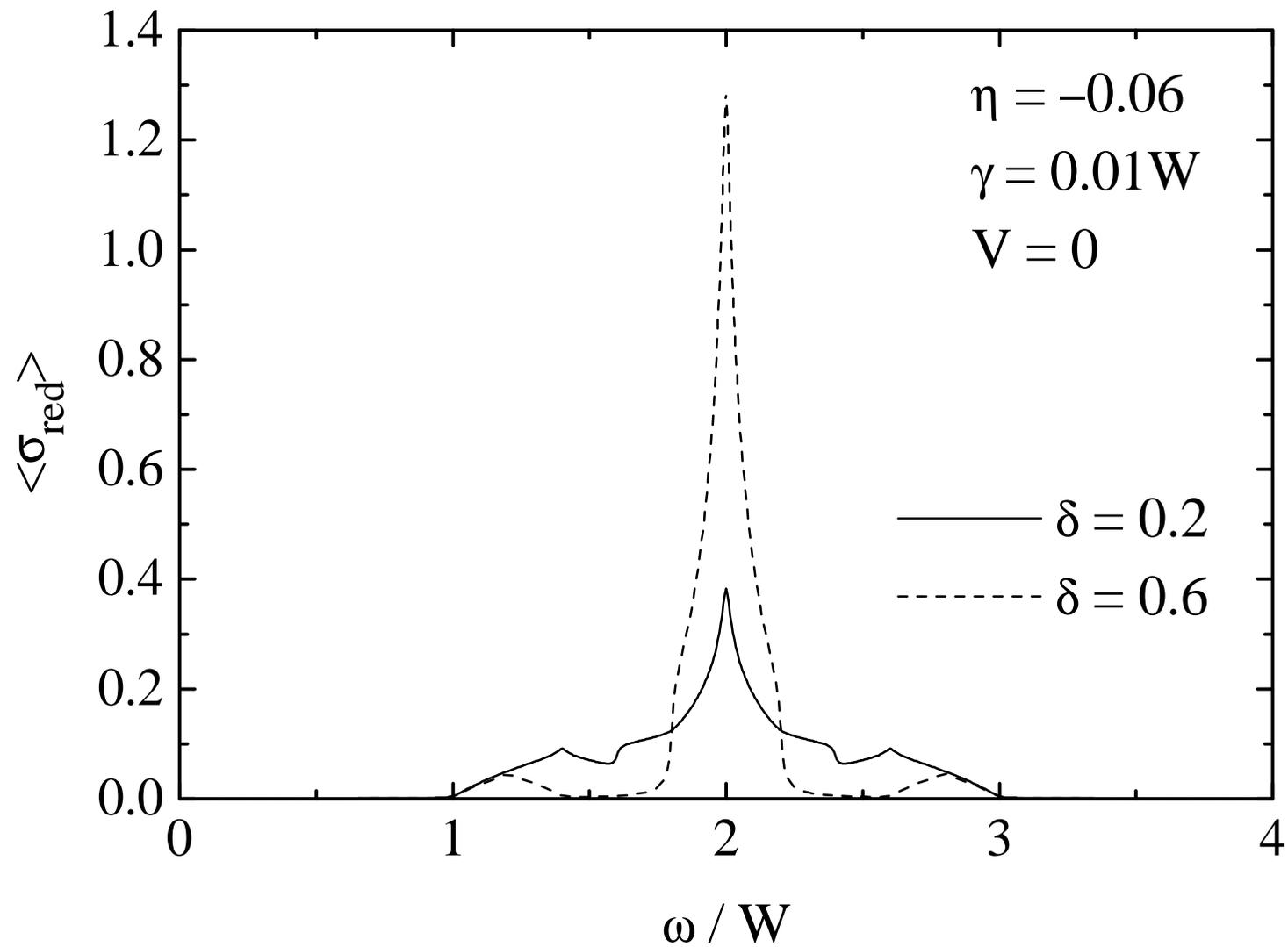
\begin{figure}[th]
\caption{Reduced average optical conductivity,
$\overline{\sigma_{\rm red}(\omega >0,\delta,\eta)}$,
in the dimerized Harris-Lange model for $U=2W$,
$\delta=0.2$ ($\delta=0.6$),
and $\eta=-0.06$. A broadening of~$\gamma=0.01W$ has been included.}
\label{hl10}
\end{figure}

For general~$\delta$, $\eta$ it is necessary to evaluate
the optical conductivity in eq.~(\ref{monstersigma}) numerically.
An example is shown in figure~\ref{hl10}.
It is seen that now there are two side-peaks in the optical absorption
spectrum.
The new peaks due to the Peierls distortion are weaker than the one
at~$\omega=U$ and vanish for $\delta\to 1$.
For large lattice distortions the dominant contribution to the side peaks
comes from the small-$q$ transitions between different Peierls
subbands. Their oscillator strength is maximum for $\omega=U\pm W\sqrt{\delta}$
which determines the position of the peaks for large $\delta$.
The Peierls gap between the bands shows up in the
optical spectrum. For small lattice distortions
all $(k,q)$ contribute. The signature of the
Peierls gap is smeared out and the position of the side peaks cannot be
expressed in terms of a simple function of~$\delta$.

\section{Optical absorption in the extended Harris-Lange model}
\label{optabsHubbard}

\subsection{Extended dimerized Harris-Lange model}

Strongly isotropic, almost ideal
one-dimensional systems like polymers or charge-transfer salts
are not properly described by the Hubbard model of eq.~(\ref{Hubb-Model})
for two reasons:
(i)~the Peierls distortion is not taken into account, and
(ii)~the residual
Coulomb interaction between charges beyond the Hubbard on-site interaction
is neglected.
The exponential decay of the Wannier wave functions naturally
allows to limit the interactions to on-site and nearest-neighbor Hubbard
terms. In the ``Zero Differential Overlap Approximation''
it is further {\em assumed\/} that only the direct Coulomb term has to be taken
into account for the nearest-neighbor Coulomb interaction
(Kivelson, Su, Schrieffer, and Heeger 1987); (Wu, Sun, and Nasu 1987);
(Baeriswyl, Horsch, and Maki 1988); (Gammel and Campbell 1988);
(Kivelson, Su, Schrieffer, and Heeger 1988); (Campbell {\em et al.} 1988);
(Painelli and Girlando 1988); (Painelli and Girlando 1989);
(Campbell, Gammel, and Loh 1990).                                      

Optical absorption spectra for the extended dimerized Hubbard model
could only be calculated numerically for small system sizes.
Within such an approach the Hamiltonian is
explicitly diagonalized
(Soos and Ramesesha 1984); (Tavan and Schulten 1986);
(Guo, Mazumdar, Dixit, Kajzar, Jarka, Kawabe, and Peyghambarian 1993);
(Guo, Guo, and Mazumdar 1994).                                           
Since the dimension of the Hilbert space increases
exponentially (${\rm dim} \hat{H} = 4^L$) the numerical analysis is
restricted to short chains ($L \leq 12$)
due to the limited computer power.

Therefore, it is natural to analytically
investigate the (dimerized) Harris-Lange model
with an additional nearest-neighbor interaction.
We will show below that the optical spectrum can still be calculated
for this model
which is equivalent to the extended dimerized Hubbard model
to order $t (t/U)^{-1}$, $t (t/U)^0$, and $t(V/U)^0$.
The dimerized extended Harris-Lange model reads
\begin{mathletters}\begin{eqnarray}
\hat{H}_{\rm HL}^{\rm dim, ext} &=&
\hat{T}_{\rm LHB}(\delta) +  \hat{T}_{\rm UHB}(\delta)
+ U \hat{D} + V \hat{V} \\[6pt]
\hat{V} &=& \sum_l (\hat{n}_l-1)(\hat{n}_{l+1}-1)
\; .
\end{eqnarray}\end{mathletters}%
For half-filling the ground state of the extended dimerized
Harris-Lange model is still $2^L$-fold spin
degenerate because every site is singly
occupied for $| V | < U/2$. The energy of these states is zero, $E_0=0$,
irrespective of the dimerization
value~$\delta$.

The double occupancy and the hole in the excited states
now experience a nearest-neighbor attraction while the spin sector
remains unchanged. Thus we may immediately translate~$\hat{V}$
into our band picture as
\begin{equation}
\hat{V}^{\rm band} = - \frac{2}{L} \sum_{|q|\leq \pi/a}\cos(qa) 
\sum_{|k|,|p|\leq \pi/a}
\hat{u}_{k+q}^+\hat{u}_{k}^{\phantom{+}}
\hat{l}_{p}^{\phantom{+}} \hat{l}_{p-q}^+
\end{equation}
which describes the scattering of a hole in the lower Hubbard band
with a particle in the upper Hubbard band. Again, the projection operators
can be disregarded for the optical absorption.

\subsection{Equation of motion technique}

In the presence of the nearest-neighbor interaction it becomes
increasingly tedious to separately calculate the exact eigenenergies
and oscillator strengths.
We rather prefer to directly calculate the optical conductivity from 
an equation of motion approach.

\subsubsection{Translational invariant case}

Since we are interested in the real part of the optical conductivity
we can concentrate on the particle current density,
\begin{equation}
\langle \hat{\jmath}_t \rangle = \frac{{\cal N}_{\perp}}{La} \left(
\langle 0(t) | \hat{\jmath} | \Psi(t)\rangle + {\rm h.c.}\right) 
\end{equation}
where $|0(t)\rangle$ and $|\Psi(t)\rangle$ are the time evolution
of the ground state with and without the external perturbation
and thus obey the corresponding Schr\"{o}dinger equations.
We have already used the fact that we want to calculate the linear absorption.

We write
\begin{eqnarray}
\langle 0(t) | \hat{\jmath} | \Psi(t)\rangle
&=& \sum_{k,q} -iea\epsilon(k)\hat{x}_{q}^+
\langle 0(t) | \hat{l}_{k-q/2}^+\hat{u}_{k+q/2}^{\phantom{+}} | \Psi(t)\rangle
\nonumber \\[3pt]
& \equiv &\sum_{k,q} \lambda(k,q) j_{k;q}(t) \; .
\end{eqnarray}
Upon Fourier transformation we obtain
\begin{equation}
\langle \hat{\jmath}_{\omega} \rangle =
\frac{{\cal N}_{\perp}}{La} \left(
\sum_{k,q} \lambda(k,q) j_{k;q}(\omega) + \lambda^+(k,q) j_{k;q}^*(-\omega)
\right) \; .
\end{equation}
Since we are interested in the optical conductivity for positive frequencies
(optical absorption) we may disregard the second term which contributes
to~$\omega < 0$. 

The equation of motion for~$j_{k;q}(t)$ becomes
\begin{mathletters}
\begin{eqnarray}
i \frac{\partial  j_{k;q}(t) }{\partial t} &=&
\langle 0(t) | \left[
\hat{l}_{k-q/2}^+\hat{u}_{k+q/2}^{\phantom{+}}, \hat{H}_{\rm HL}^{\rm band}
\right]_{-} | \Psi(t)\rangle
- \frac{{\cal A}(t)}{c} 
\langle 0(t) |
\hat{l}_{k-q/2}^+\hat{u}_{k+q/2}^{\phantom{+}}
\hat{\jmath} | 0(t)\rangle
 \\[6pt]
\omega j_{k;q}(\omega) &=& E(k,q) j_{k;q}(\omega)
- 2V\left( \cos(ka) j_q^c(\omega) +\sin(ka)j_q^s(\omega)\right)
- \lambda^+(k,q) \frac{{\cal A}(\omega)}{c} 
\end{eqnarray}
\end{mathletters}%
where we kept the expansion linear in the external perturbation,
and performed the Fourier transformation.
Furthermore, we introduced the abbreviations
\begin{equation}
j_q^{c,s}(\omega) = \frac{1}{L}
\sum_k \left( {\cos (ka) \atop \sin (ka)}\right) j_{k;q}(\omega) \; .
\end{equation}
For our calculations we only need~$j_q^c(\omega)$ since our current
operator preserves parity.
The particle current density for positive frequencies becomes
\begin{equation}
\langle \hat{\jmath}_{\omega>0}\rangle = \frac{{\cal N}_{\perp}}{a}(2tiea)
\sum_q \hat{x}_q^+ j_{q}^c(\omega) 
\end{equation}
which is proportional to the external field.

We introduce the function
\begin{equation}
F(q) = \frac{2}{L} \sum_{|k|\leq \pi/a} \frac{(\cos ka)^2}{\omega - E(k,q)}
\end{equation}
which allows us to finally express the optical conductivity as
\begin{equation}
{\rm Re} \{ \overline{\sigma(\omega>0,V)} \}
= - \frac{(Wea)^2{\cal N}_{\perp}}{16 a\omega} \frac{1}{L}
\sum_{|q|\leq \pi/a} g(q) {\rm Im}\left\{ \frac{F(q)}{1+VF(q)} \right\} \; .
\label{ResigmaHLV}
\end{equation}
The result will be discussed in the next subsection.

\subsubsection{Extended dimerized Harris-Lange model}

The same procedure can be applied to the dimerized case where it is best
to start from the diagonalized Hamiltonian in the form of
eq.~(\ref{effHL}),
and the current operator in the form of eq.~(\ref{jinterdimHL}).
The calculations are outlined in appendix~\ref{appc}.

We introduce the three functions $F_{1,2,3}$ as
\begin{mathletters}
\label{capitalF}
\begin{eqnarray}
F_{1} (q) &=& \frac{2}{L} \sum_{|k| \leq \pi/(2a)}
\cos^2(ka)  \Biggl[ |f_{+,+}|^2 
\left(  \frac{1}{\omega-E_{-,-}} +\frac{1}{\omega-E_{+,+}} \right)
\nonumber \\[3pt]
&& \phantom{\frac{2}{L} \sum_{|k| \leq \pi/(2a)}
\cos^2(ka)  \Biggl[ }
+|f_{+,-}|^2
\left(\frac{1}{\omega-E_{-,+}}+\frac{1}{\omega-E_{+,-}}\right)
\Biggr] \\[6pt]
F_{2} (q) &=& \frac{2}{L} \sum_{|k| \leq \pi/(2a)}
\sin^2(ka)  \Biggl[ |f_{+,-}|^2 
\left(  \frac{1}{\omega-E_{-,-}} +\frac{1}{\omega-E_{+,+}} \right)
\nonumber \\[3pt]
&& \phantom{\frac{2}{L} \sum_{|k| \leq \pi/(2a)}
\sin^2(ka)  \Biggl[ }
+|f_{+,+}|^2
\left(\frac{1}{\omega-E_{-,+}}+\frac{1}{\omega-E_{+,-}}\right)
\Biggr]
\\[6pt]
F_{3} (q) &=& \frac{2}{L} \sum_{|k| \leq \pi/(2a)}
\cos(ka)\sin(ka) \Biggl\{ f_{+,+}^{\phantom{*}} f_{+,-}^* \left[
\frac{1}{\omega-E_{+,+}}+ \frac{1}{\omega-E_{-,-}} \right]
\nonumber \\[3pt]
&& \phantom{\frac{2}{L} \sum_{|k| \leq \pi/(2a)}
\cos(ka)\sin(ka) \biggl\{  }
- f_{+,+}^* f_{+,-}^{\phantom{*}} 
\left[  \frac{1}{\omega-E_{-,+}}+ \frac{1}{\omega-E_{+,-}} \right]\Biggr\}
\end{eqnarray}
\end{mathletters}%
where $f_{\tau,\tau'}\equiv f_{\tau,\tau'}(k,q)$
and $E_{\tau,\tau'}=E_{\tau,\tau'}(k,q)$
were introduced in eq.~(\ref{thefs}) and eq.~(\ref{Etautauprime}).
Furthermore, we abbreviate~$A_j=(1+\delta\eta)F_j-(\eta+\delta)F_3$,
$B_j=(\delta+\eta)F_j-(1+\delta\eta)F_3$ ($j=1,2$),
$C_1=(1+\delta\eta)^2 F_1+(\delta+\eta)^2 F_2 
-2(1+\delta\eta)(\delta+\eta)F_3$, and
$C_2=(1+\delta\eta)^2 F_2+(\delta+\eta)^2 F_1 
-2(1+\delta\eta)(\delta+\eta)F_3$.
The real part of the average optical conductivity can then be expressed as
\begin{eqnarray}
{\rm Re}\{\overline{\sigma(\omega >0, V, \delta,\eta)} \}
&=&
{\rm Re}\{\overline{\sigma(\omega >0,\delta,\eta)} \}
\nonumber \\[6pt]
&&
+ \frac{V{\cal N}_{\perp}(Wea)^2}{16 a\omega L} {\rm Im}\Biggl\{
\sum_{|q| \leq \pi/(2a)}
\frac{1}{(1+VF_1)(1+VF_2)-(VF_3)^2}
\label{thefinalresultHL}
\\[6pt]
&& 
\phantom{+ \frac{V{\cal N}_{\perp}(Wea)^2}{16 a\omega L} {\rm Im}\Biggl\{ }
\biggl\{
g(q) \left[ A_1^2 +B_2^2 +V(F_1F_2-F_3^2) C_1 \right]  \nonumber \\
&& \phantom{+ \frac{V{\cal N}_{\perp}(Wea)^2}{16 a\omega L} {\rm Im}\Biggl\{
\biggl\{ }
+g(q+\frac{\pi}{a}) \left[ A_2^2 +B_1^2 +V(F_1F_2-F_3^2) C_2 \right] 
\biggr\} \Biggr\} \; .
\nonumber
\end{eqnarray}
The result for~$V=0$ is given in eq.~(\ref{monstersigma}).
In the following
we will discuss the results for the average optical absorption in the
presence of a nearest-neighbor interaction.

\subsection{Optical absorption}

\subsubsection{Translational invariant case}

The help function~$F(q)$ can be calculated analytically
with the help of eqs.~(2.267,1), (2.266), and~(2.261) 
of (Gradshteyhn and Ryzhik 1980).               
The result is
\begin{equation}
F(q) \! = \! \frac{2}{[4t\sin(qa/2)]^2}
\! \left\{
\!
\begin{array}{lcl}
\omega-U -\sqrt{(\omega-U)^2- \left(4t\sin(qa/2)\right)^2\, }
& {\rm for} & |\omega-U| \geq |4t\sin(qa/2)|
\\[6pt]
-i \sqrt{ \left(4t\sin(qa/2)\right)^2 - (\omega-U)^2\, }
& {\rm for} & |\omega-U| < |4t\sin(qa/2)|
\end{array}
\right.
\, .
\end{equation}
The result for~$V=0$, eq.~(\ref{what}),
follows after the substitution of~$x=\sin(qa/2)$ into
equation~(\ref{ResigmaHLV}).

The total optical absorption is shown in figure~\ref{hl01}.
For arbitrarily small~$V>0$ there is a bound exciton which
is the standard situation for one-dimensional short-range attractive
potentials between a positive (hole) and negative charge (double occupancy).
This is evident from the form of~$F(q)$ which allows for excitons
with momenta~$q$ if $|\omega -U| \geq |4t\sin(qa/2)|$
which is fulfilled for~$q=0$ for all~$V>0$.

When the attraction between the two opposite charges is strong,
the full exciton band with width~$W_{\rm exc}=4t^2/V$ is formed.
This can be seen from the zeros of the denominator in eq.~(\ref{ResigmaHLV})
in the region~$|\omega-(U-V-W_{\rm exc}/2)|\leq W_{\rm exc}/2$.
One finds from $1+VF(q)=0$ that
\begin{eqnarray}
\omega= U-V-\frac{W_{\rm exc}}{2} + \frac{W_{\rm exc}}{2} \cos qa \; .
\end{eqnarray}
This is precisely the dispersion relation for bound pairs in one dimension
with nearest-neighbor hopping of strength~$t_{\rm exc}=t (t/V)$:
at large~$V$ the excitons are essentially nearest-neighbor pairs
of opposite charges which {\em coherently\/} move with the hopping
amplitude~$t_{\rm exc}$. Note that this motion requires an intermediate
(``virtual'') configuration where the two charges are not nearest neighbors.
Consequently, the hopping integral of the individual constituents,~$t$,
is reduced by the factor~$t/V$ for the motion of the pair.
The full band becomes apparent when $W_{\rm exc}+V > W$ or $V > W/2$,
see figure~\ref{hl01}. Recall that the spin sector provides {\em any\/}
momentum to the charge sector. The momentum transfer, however, is
modulated by the function~$g(q)$ which is maximum at $q=\pi/a$
and reflects the fact that states with antiferromagnetic spin correlations
are best suited for optical absorptions since they contain many neighboring
singlet pairs. Hence, the $q=\pi/a$-exciton dominates over
the $q=0$-exciton for $V>W/2$.

It is amusing to see that the optical absorption of a Peierls insulator
and a Mott-insulator (extended Harris-Lange model: $U \gg W$, $V> W/2$,
$J=0$) can look very similar,
compare figure~2 of~I and figure~\ref{hl01}.
This has already been noted long time ago by Simpson
(Simpson 1951); (Simpson 1955); (Salem 1966); (Fave 1992)           
who explained the optical absorption spectra of short polyenes in the
above exciton model.
It is seen that Simpson's model is naturally included in our strong-correlation
approach.
For real polymers, however, Simpson's original approach is not satisfactory.
A fully developed exciton band
only exists in the presence of an
incoherent spin background. Now that
even the spin-Peierls effect is excluded one can by no means
explain the Peierls distortion of the lattice
as an electronic effect.
This does not exclude other, e.g., extrinsic, explanations for a
lattice distortion.

\subsubsection{Extended dimerized Harris-Lange model}

The full spectrum has to be determined numerically. An example
for various values of~$V/t$ is shown in figure~\ref{hl11}.
For large~$V/t$ we have a fully developed
exciton band
which is itself Peierls-split into two branches.
Thus one obtains four van-Hove singularities in the optical
absorption spectrum.
The Peierls gap is given by~$\Delta_{\rm exc}^{\rm P}=
\delta W_{\rm exc}= 4t^2\delta/V$.
Even for~$V=W$ it is smeared out since $V/t$ is not too large yet
and the phenomenological damping~$\gamma$ is already of the order
of the gap.

For small~$V/t$ we obtain the signature
of the $q=0$ exciton for~$\delta=0$, $V>0$.
For intermediate~$V$ this peak develops into a van-Hove singularity
of the upper exciton subband. The signatures of the second
van-Hove singularity of the upper band are clearly visible for~$V=W/2$.
The peaks of the Peierls subbands for~$V=0$, $\delta \neq 0$
are both red-shifted. The peak at 
lower energy increases in intensity and finally forms
the lower exciton subband while the peak at higher energy
quickly looses its oscillator strength.

\section{Summary and Outlook}

In this paper we addressed the optical absorption of
the half-filled Harris-Lange model which
is equivalent to the Hubbard model at strong correlations
and temperatures large compared to the spin energy scale. 
It is extremely difficult to analytically
calculate optical properties of interacting electrons in one dimension.
For strong coupling when the Hubbard interaction is large compared
to the band-width matters considerably simplified since
the energy scales for the charge and spin excitations are well separated.
We were able to derive an {\em exactly\/} equivalent
band structure picture for the charge degrees
of freedom and found that the upper and lower Hubbard band are actually
{\em parallel\/} bands with the band structure of free Fermions.
We have taken special
care of the spin background which can act as a momentum
reservoir for the charge system. 
Since we can exactly integrate out the
spin degrees of freedom for the Harris-Lange model
we were able to solve the problem even in the presence
of a lattice dimerization and a nearest-neighbor interaction between
the electrons.

For a vanishing nearest-neighbor interaction
we found a prominent absorption peak at~$\omega=U$ and additional side peaks
in the absorption bands $|\omega-U| \leq W$
in the presence of a lattice distortion~$\delta$.
When a further nearest-neighbor interaction between the charges
was included, we found the formation of
Simpson's exciton band of band-width $W_{\rm exc.}=4t^2/V$ for $V>W/2$
which is eventually Peierls-split.
As usual the excitons draw almost all oscillator strength from the
band.

It should be clear that the Harris-Lange model with its highly degenerate
ground state is not a suitable model for the study of real materials.
The results presented here are relevant to systems for
which the temperature is much larger than the spin exchange energy.
Real experiments are not carried out in this ``hot-spin'' regime
but at much lower temperatures for which the system
is in an unique ground state with antiferromagnetic correlations. 
Unfortunately, this problem cannot be solved analytically.
In the third and last paper of this series 
(Gebhard {\em et al.} III 1996) we will               
employ the analogy to an ordinary semiconductor (electrons and holes
in a phonon bath) to design a ``no-recoil'' approximation 
for the chargeons in a spinon bath.
It will allows us to determine the coherent absorption features 
of the Hubbard model at large~$U/t$.

\section*{Acknowledgments}

We thank H.~B\"{a}\ss ler, A.~Horv\'{a}th,
M.~Lindberg, S.~Mazumdar, M.~Schott, and
G.~Weiser for useful discussions.
The project was supported in part by the
Sonderforschungsbereich~383 
``Unordnung in Festk\"{o}rpern
auf mesoskopischen Skalen'' of the Deutsche Forschungsgemeinschaft.

\newpage
\begin{appendix}

\section{The Harris-Lange model}
\label{appeigen}

\subsection{Sum rules}
\label{appsumrule}

We briefly account for the sum rules. We have
\begin{equation}
\int_{0}^{\infty} d\omega\ {\rm Im}\{\chi(\omega)\}
= \pi \frac{{\cal N}_{\perp}}{La} \sum_n \left| \langle 0 |
\hat{\jmath}^2|n\rangle\right|^2 = \pi \frac{{\cal N}_{\perp}}{La}
\langle 0 | \hat{\jmath}^2|0\rangle \; .
\end{equation}
It is a standard exercise to show that
\begin{equation}
\langle 0 | \hat{\jmath}^2 | 0 \rangle =
(2tea)^2 \sum_l \left(1+(-1)^l\delta\right)^2\left(1+(-1)^l\eta\right)^2
\langle 0 | \left( \frac{1}{4}
-\hat{\rm\bf S}_l\hat{\rm\bf S}_{l+1}\right) | 0 \rangle \; .
\end{equation}
We define the positive quantities
\begin{equation}
C_S^{\rm even, odd}= \frac{1}{L} \sum_l \frac{1\pm (-1)^l}{2}
\langle 0 | \left( \frac{1}{4}
-\hat{\rm\bf S}_l\hat{\rm\bf S}_{l+1}\right) | 0 \rangle
\label{CSevenCSodd}
\end{equation}
and may then write
\begin{equation}
\int_{0}^{\infty} d\omega\ {\rm Im}\{\chi(\omega)\}
= \pi {\cal N}_{\perp}a (2te)^2 \left[
\left(1+\delta\right)^2\left(1+\eta\right)^2
C_S^{\rm even}
+
\left(1-\delta\right)^2\left(1-\eta\right)^2
C_S^{\rm odd}
\right] \; .
\end{equation}
If we average over all possible states~$|0\rangle$ we obtain
$C_S^{\rm even, odd}=1/8$ since only the singlet configuration contributes.
Hence,
\begin{equation}
\int_{0}^{\infty} d\omega \overline{{\rm Im}\{\chi(\omega)\}}
= 
\pi {\cal N}_{\perp}a (te)^2 \left[(1+\delta\eta)^2 +(\delta+\eta)^2\right]
\; .
\end{equation}
The area under the curves for~$\sigma_{\rm red}(\omega)$, eq.~(\ref{sigmared}), 
are thus given by
\begin{mathletters}
\begin{eqnarray}
\int_0^{\infty} \frac{d\omega}{W} \sigma_{\rm red}(\omega) &=&
\frac{\pi}{4}
\left[
\left(1+\delta\right)^2\left(1+\eta\right)^2 C_S^{\rm even}
+
\left(1-\delta\right)^2\left(1-\eta\right)^2 C_S^{\rm odd}
\right]
\label{sumrulesigmared}
\\[6pt]
\int_0^{\infty} \frac{d\omega}{W} \overline{\sigma_{\rm red}(\omega)} &=&
\frac{\pi}{16} \left[(1+\delta\eta)^2 +(\delta+\eta)^2\right]
\; .
\end{eqnarray}
\end{mathletters}%

\subsection{Momentum of eigenstates}
\label{appmomentum}
First we will assume that the number of sites and the
number of particles is even. $N_C$ and~$N_S$ will thus also be even.
Let~$\hat{{\cal T}}$ be the translation operator by one site,
i.~e., $\hat{{\cal T}} \hat{c}_{l,\sigma}\hat{{\cal T}}^{-1}=
\hat{c}_{l+1,\sigma}$.
We have to show that
\begin{equation}
\hat{{\cal T}} |\Psi\rangle = e^{-iPa}|\Psi\rangle
\end{equation}
holds. We have to distinguish two cases: (i)~a spin is at site~$L$
in $|\Psi\rangle$ or (ii)~a charge is at site~$L$
in $|\Psi\rangle$.

\paragraph{case~(i):}
the operator~$\hat{{\cal T}}$ shifts the spin from site~$L$ to the first site.
This results in a phase factor~$(-1)$ since one has to commute the Fermion
operator~$(N-1)$-times to obtain the proper order in~$|\Psi\rangle$.
Furthermore, the states in~$|\Psi\rangle$ have, relative to those in
$\hat{{\cal T}}|\Psi\rangle$,
\begin{enumerate}
\item shifted the spin sequence by one unit. This results in a phase factor
$\exp(ik_Sa)$;
\item 
an additional factor $(-1)$ for each doubly occupied site.
This gives a phase factor~$\exp(i\pi N_d)$;
\item
a Slater determinant in which each site index is shifted by one.
This results in a phase factor
$\exp(i\sum_{j=1}^{N_C} (k_j+\Phi_{CS})a)$.
\end{enumerate}
In sum we obtain
\begin{eqnarray}
(-1) &=& e^{-iPa} e^{ik_Sa} e^{i\pi N_d} e^{i\sum_{j=1}^{N_C} (k_j+\Phi_{CS})a}
\nonumber \\[6pt]
P&=& k_S +(\pi/a) (N_d-1)+ \sum_{j=1}^{N_C} (k_j+\Phi_{CS})
 \quad {\rm mod\ } 2\pi/a \; .
\end{eqnarray}

\paragraph{case~(ii):}
the operator~$\hat{{\cal T}}$ shifts the charge from site~$L$ to the first
site.
The states in~$|\Psi\rangle$ have, relative to those in
$\hat{{\cal T}}|\Psi\rangle$,
\begin{enumerate}
\item shifted the charge sequence by one unit. This results in a phase factor
$\exp(ik_Ca)$;
\item 
an additional factor $(-1)$ for each doubly occupied site.
This gives a phase factor~$\exp(i\pi N_d)$;
\item
a Slater determinant in which
\begin{enumerate}
\item each site index is shifted by one.
This results in a phase factor
$\exp(i\sum_{j=1}^{N_C} (k_j+\Phi_{CS})a)$;
\item
the last row and the first row are interchanged.
This gives an additional factor $(-1)^{N_C-1}=-1$;
\item
the first row is $(1,\ldots 1)$ instead of
$\left(\exp(i(k_1+\Phi_{CS})La),\ldots\exp(i(k_{N_C}+\Phi_{CS})La)\right)$
$= (1,\ldots 1)\exp(i(k_C-k_C)a)$.
This gives an additional phase factor~$\exp(-i(k_C-k_S)a)$.
\end{enumerate}
\end{enumerate}
In sum we obtain
\begin{eqnarray}
1 &=& e^{-iPa} e^{ik_C a} e^{i\pi N_d} e^{i\sum_{j=1}^{N_C} (k_j+\Phi_{CS})a}
(-1) e^{-i(k_C-k_S)a}
\nonumber \\[6pt]
P&=& k_S +(\pi/a) (N_d-1)+ \sum_{j=1}^{N_C} (k_j+\Phi_{CS}) \quad {\rm mod\ }
2\pi/a
\end{eqnarray}
as before. If the number of particles~$N$ is odd, the momentum is shifted
by another factor of~$\pi/a$, if one repeats the above arguments.
This proves that the total momentum of the state~$|\Psi\rangle$
is indeed given by eq.~(\ref{eigenmomHL}).

\subsection{Energy of eigenstates}
\label{appenerg}
We want to prove that
\begin{equation}
\hat{H}_{\rm HL} |\Psi\rangle = E|\Psi\rangle
\quad ; \quad 
E = \sum_{j=1}^{N_C} \epsilon(k_j+\Phi_{CS}) +U N_d  \; .
\label{showit}
\end{equation}
Again we restrict ourselves to even~$N$.
The bulk terms are simple since there is no hopping across the boundary.
A hopping process of a double occupancy and a hole
are equivalent:
\begin{mathletters}
\begin{eqnarray}
(-1)^l | \ldots \bullet_l \sigma_{l+1} \ldots\rangle &\mapsto&
- (-1)^l | \ldots \sigma_{l} \bullet_{l+1}  \ldots\rangle
= (-1)^{l+1} | \ldots \sigma_{l} \bullet_{l+1}  \ldots\rangle \\[3pt]
| \ldots \circ_l \sigma_{l+1} \ldots\rangle &\mapsto&
| \ldots \sigma_{l} \circ_{l+1}  \ldots\rangle \; .
\end{eqnarray}
\end{mathletters}%
The extra minus sign which appears when a double occupancy moves
has been taken care of in the wave function
by the phase factor~$(-1)^l$ for a double
occupancy at site~$l$. Now that there is no difference in the motion of
double occupied sites and holes they dynamically behave as
spinless Fermions. Since the Slater determinant
is the proper phase factor for non-interacting Fermions
eq.~(\ref{showit}) holds for the bulk terms.
This also shows that only the Harris-Lange model with
hopping amplitudes~$|t_{\rm LHB}|=|t_{\rm UHB}|$ can be
solved. Another integrable but trivial
case is~$t_{\rm LHB}=0$, $t_{\rm UHB}\neq 0$
and vice versa.

We now address the boundary terms.
A typical configuration for which transport across the boundary is
possible is $|S_1, \ldots C_L\rangle$.
The phase of the configuration is given by a Slater determinant
in which the last row has the entry
$\left(\exp(i(k_1+\Phi_{CS})La),\ldots\exp(i(k_{N_C}+\Phi_{CS})La)\right)$
$= (1,\ldots 1)\exp(i(k_C-k_S)La)$.
The action of~$\hat{H}_{\rm HL}$ moves the charge from site~$L$
to the first position by which an extra minus sign occurs
since the electron operator for the spin
had to be commuted with~$(N-1)$ other electron operators.
These phase factors have to be compared to the corresponding
configuration in~$E|\Psi\rangle$.
Relative to the configuration in~$\hat{H}_{\rm HL}|\Psi\rangle$
it has
\begin{enumerate}
\item shifted the spin sequence by minus one unit.
This results in a phase factor~$\exp(-ik_Sa)$;
\item shifted the charge sequence by one unit.
This results in a phase factor~$\exp(ik_Ca)$;
\item a Slater determinant which has
$\left(\exp(i(k_1+\Phi_{CS})a),\ldots\exp(i(k_{N_C}+\Phi_{CS})a)\right)$
in the first row.
\end{enumerate}
The boundary terms should give the same result as the bulk terms.
This leads to the condition
\begin{equation}
- e^{i(k_C-k_S)La} = e^{-ik_Sa} e^{ik_C a} (-1)^{N_C-1}
\end{equation}
which is obviously fulfilled. The proof for odd~$N$ is analogous,
and eq.~(\ref{showit}) holds for all~$N$.

\subsection{Electrical dipole operator for the Harris-Lange model}
\label{appdipoleHL}

We can derive the electrical
dipole operator for the Harris-Lange model from its definition in
equation~(A.22) of~I. We use the Hamilton operator in the band picture
interpretation, eq.~(\ref{effHL}), and the corresponding current operator,
eq.~(\ref{jhlbandpicture}). Since~$\hat{x}_q$ can be replaced by its
average value~$\sqrt{g(q)/(2L)}$ one easily sees that
the dipole operator becomes
\begin{mathletters}
\begin{eqnarray}
\hat{P}_{\rm inter}^{\rm HL}&=&\sum_{|k|,|q|\leq \pi/a}
\mu_{\rm inter}^{\rm HL}(k,q)
\left( \hat{u}_{k+q/2}^+\hat{l}_{k-q/2}^{\phantom{+}} + \hat{l}_{k-q/2}^+
\hat{u}_{k+q/2}^{\phantom{+}} \right)
\\[3pt]
\mu_{\rm inter}^{\rm HL}(k,q)&=& i \frac{\lambda(k,q)}{E(k,q)}=
ea \sqrt{\frac{g(q)}{2L}} \frac{\epsilon(k)}{E(k,q)}
\end{eqnarray}
\end{mathletters}%
with~$E(k,q)=U+\epsilon(k+q/2)-\epsilon(k-q/2)$.
One sees that the dipole matrix element is of the order~$t/U$ as it should
be for interband transitions.
Furthermore, for small momentum transfer we obtain
\begin{equation}
\mu_{\rm inter}^{\rm HL}(k;q\to 0)  \sim \epsilon(k) \; .
\end{equation}
This is the correct form since we create a neighboring
hole and double occupancy which corresponds to an electric dipole between
nearest neighbors.

The procedure is readily generalized to the dimerized Harris-Lange model.
The interband current operator in terms of the Fermion operators
for the four Peierls subbands is given in eq.~(\ref{jinterdimHL}),
and the diagonalized Hamiltonian can be found in eq.~(\ref{bandhldim}).
One readily finds
\begin{mathletters}
\begin{eqnarray}
\hat{P}_{\rm inter}^{\rm dim.\ HL}&=& \sum_{\tau,\tau'=\pm 1}
\sum_{|k|,|q|\leq \pi/(2a)}
\mu_{\rm inter;\tau,\tau'}^{\rm dim.\ HL}(k,q)
\left( \hat{u}_{k+q/2,\tau'}^+\hat{l}_{k-q/2,\tau}^{\phantom{+}} +
\hat{l}_{k-q/2,\tau}^+\hat{u}_{k+q/2,\tau'}^{\phantom{+}} \right)
\\[3pt]
\mu_{\rm inter;\tau,\tau'}^{\rm HL}(k,q)&=&
i \frac{\lambda_{\tau,\tau'}(k,q)}{E_{\tau,\tau'}(k,q)}
\end{eqnarray}
\end{mathletters}%
with $\lambda_{\tau,\tau'}(k,q)$ as the root
of eq.~(\ref{thelambdassquared}), and
$E_{\tau,\tau'}(k,q)=U+\tau' E(k+q/2)-\tau E(k-q/2)$, see
eq.~(\ref{Etautauprime}).

The dipole matrix elements again simplify for small~$q$.
Note that one obtains both contributions from $q\to 0$ and $q\to \pi/a$.
After some calculations one obtains
\begin{mathletters}
\label{lambdatautauprimeqzero}
\begin{eqnarray}
\left|\lambda_{+,+}(k;q=0)\right|^2 &=& \frac{1}{2L} \Biggl\{
\frac{1}{3} \left[ea\left( E(k)+\delta\eta\frac{(2t)^2}{E(k)}\right)
\right]^2 \nonumber \\[6pt]
&& \phantom{ \frac{1}{4L} \biggl[ }
+ 3 \left[ea \left(\eta E(k)-\delta\frac{(2t)^2}{E(k)}\right)
\right]^2 \Biggr\} \\[6pt]
\left|\lambda_{+,-}(k;q= 0)\right|^2 &=& \frac{1}{2L}
\left( ea \frac{\epsilon(k)\Delta(k)(1-\delta^2)}{\delta E(k)}\right)^2
\left( \frac{\eta^2}{3} +3\right)
\; .
\end{eqnarray}
\end{mathletters}%
Note that the dipole matrix elements $\left|\lambda_{+,-}(k;q=0)\right|^2$
contain the contributions from~$q= \pi/a$ for $\delta=\eta=0$.
Eqs.~(\ref{lambdatautauprimeqzero})
have to be compared to the corresponding expressions for the
Peierls chain. It is seen that the expressions display some similarities
but they show subtle differences. Even for~$q=0$ the
expressions~(\ref{lambdatautauprimeqzero}) could not have been guessed.

The corresponding dipole matrix elements become
\begin{mathletters}
\label{mutautauprimeqzero}
\begin{eqnarray}
\left|\mu_{+,+}(k;q\to 0)\right|^2 &=& \frac{1}{U^2}
\left|\lambda_{+,+}(k;q\to 0)\right|^2\\[6pt]
\left|\mu_{+,-}(k;q\to 0)\right|^2 &=& \frac{1}{(U-2E(k))^2}
\left|\lambda_{+,-}(k;q\to 0)\right|^2 \\[6pt]
\left|\mu_{-,+}(k;q\to 0)\right|^2 &=& \frac{1}{(U+2E(k))^2}
\left|\lambda_{+,-}(k;q\to 0)\right|^2 \; .
\end{eqnarray}
\end{mathletters}%
The dipole matrix elements between the same Peierls subbands are
always strong, irrespective of~$k$ or $\delta$.
However, the dipole matrix elements for transitions between different
subbands are small for strong dimerization.
Furthermore, they are small in the vicinity of
the center and the edge of the reduced Brillouin zone.

\subsection{Spin average in the Harris-Lange model}
\label{spiav}
We need to calculate
\begin{equation}
\langle \hat{x}_q^+\hat{x}_{q'}\rangle =
\frac{1}{2^L}\sum_{|0\rangle} \hat{x}_q^+\hat{x}_{q'}
\end{equation}
where
{\arraycolsep=0pt\begin{eqnarray}
\hat{x}_q^{+}(\delta,\eta)\hat{x}_{q'}^{\phantom{+}}(\delta,\eta)
&=& 
\sum_{S_1^{\prime},\ldots S_{L-2}^{\prime}}
\frac{1}{L^2} \sum_{l,r} e^{i(ql-q'r)a}
\bigl(1+\eta\delta +(-1)^l(\delta+\eta)\bigr)
\bigl(1+\eta\delta +(-1)^r(\delta+\eta)\bigr)
\nonumber \\[6pt]
&& \phantom{\sum_{S_1^{\prime},\ldots S_{L-2}^{\prime}}
\frac{1}{L^2} \sum_{l,r} e^{i(ql-q'r)} }
\langle 0 | S_1^{\prime},\ldots S_{l-1}^{\prime},
\left( \uparrow_{l}\downarrow_{l+1}-\downarrow_{l}\uparrow_{l+1}\right),
S_{l}^{\prime},\ldots S_{L-2}^{\prime}\rangle \nonumber
\\[6pt]
&& \phantom{\sum_{S_1^{\prime},\ldots S_{L-2}^{\prime}}
\frac{1}{L^2} \sum_{l,r} e^{i(ql-q'r)} }
\langle S_{L-2}^{\prime},\ldots S_{r}^{\prime},
\left( \downarrow_{r+1} \uparrow_{r}-\uparrow_{r+1}\downarrow_{r}\right),
S_{r-1}^{\prime},\ldots S_{1}^{\prime} | 0 \rangle \; .
\nonumber \\
&&  \label{xqxqprimeapp} 
\end{eqnarray}}%
Since the set of spin states $|0\rangle$ is complete we
may exactly trace it out and are left with the
calculation of the spin matrix elements
\begin{eqnarray}
 &&M(l,r)=\frac{1}{2^L}\sum_{S_1^{\prime},\ldots S_{L-2}^{\prime}}
 \\[3pt]
 && \langle S_{L-2}^{\prime},\ldots S_{r}^{\prime},
\left( \downarrow_{r+1} \uparrow_{r}-\uparrow_{r+1}\downarrow_{r}\right),
S_{r-1}^{\prime},\ldots S_{1}^{\prime}  
|
S_1^{\prime},\ldots S_{l-1}^{\prime},
\left( \uparrow_{l}\downarrow_{l+1}-\downarrow_{l}\uparrow_{l+1}\right),
S_{l}^{\prime},\ldots S_{L-2}^{\prime} \rangle \; . \nonumber
\end{eqnarray}%
The value of all spins between the sites~$l$ and~$r$ is fixed by
the singlet operators at $(l, l+1)$ and  $(r, r+1)$.
We find
\begin{equation}
M(l,r) = 2(-1)^{r-l} 2^{-|r-l|-2}
\end{equation}
which shows that the correlation function for finding
two singlet pairs at distance~$n=|r-l|$ exponentially decays 
with correlation length~$\xi_{\rm S}=1/\ln(2)$.

Since $M(l,r)$ only depends on the distance between the two sites
we may carry out one of the lattice sums in 
equation~(\ref{xqxqprimeapp}). This gives
\begin{eqnarray}
\langle \hat{x}_q^+\hat{x}_{q'}\rangle =
\frac{1}{2L}\biggl\{ 
&&
 \delta_{q,q'} \sum_{n=-L/2}^{L/2-1}
e^{inqa}2^{-|n|}(-1)^n \left[(1+\delta\eta)^2+(-1)^n(\delta+\eta)^2\right]
 \\[3pt] \nonumber
&& + \delta_{q,q'+\frac{\pi}{a}} \sum_{n=-L/2}^{L/2-1}
e^{inqa}2^{-|n|} (1+\delta\eta)(\delta+\eta)(1+(-1)^n)\biggr\} \; .
\end{eqnarray}
The sum over~$n$ is readily taken and gives the final result
\begin{mathletters}
\begin{eqnarray}
\langle \hat{x}_q^+\hat{x}_{q'}\rangle &=&
\frac{1}{2L} \biggl\{ 
\delta_{q,q'} 
\left[(1+\eta\delta)^2 g(q)+(\delta+\eta)^2 g(q+\frac{\pi}{a})\right]
\nonumber \\[3pt]
&& \phantom{\frac{1}{2L} \biggl\{ }
+ \delta_{q,q'+\frac{\pi}{a}} (1+\eta\delta)(\delta+\eta) 
\left[g(q)+ g(q+\frac{\pi}{a})\right]
\biggr\}
\end{eqnarray}
with the help function
\begin{equation}
g(q) = \frac{3}{5+4\cos(qa)} \; .
\end{equation}\end{mathletters}%
In particular, for the translational invariant case we find
\begin{equation}
\langle \hat{x}_q^+\hat{x}_{q'}\rangle =
\delta_{q,q'}\frac{g(q)}{2L} \; .
\end{equation}
This shows that $\hat{x}_q$ can be replaced by $\sqrt{g(q)/(2L)}$
in the translational invariant case.

\section{Equation of motion technique for the extended dimerized
Harris-Lange model}
\label{appc}

Here we briefly outline the calculations for the most general case~$V\neq 0$,
$\delta\neq 0$.
We use the diagonalized band picture Hamiltonian in the form of
eq.~(\ref{effHL}),
and the band picture current operator in the form of eq.~(\ref{jinterdimHL}).
This has the advantage that the equation of motion can directly be inverted
and the contribution for~$V=0$ can immediately be separated.

The equations of motions give for the four currents
\begin{equation}
j_{\tau,\tau'}(k,q;\omega) = -\frac{
({\cal A}(\omega)/c)\lambda_{\tau,\tau'}^+
+ 2V
\left( \cos(ka) X_{\tau,\tau'}^c +\sin(ka) X_{\tau,\tau'}^s
\right)
}
{\omega- E_{\tau,\tau'}}
\label{C1}
\end{equation}
where~$E_{\tau,\tau'}\equiv E_{\tau,\tau'}(k,q)=
U+\tau' E(k+q/2)-\tau E(k-q/2)$,
$\lambda_{\tau,\tau'}\equiv\lambda_{\tau,\tau'}(k,q)$, and
$X_{\tau,\tau'}^{c,s}\equiv X_{\tau,\tau'}^{c,s}(k,q)$.
The difficult terms~$X^{c,s}\equiv X_{+,+}^{c,s}=-X_{-,-}^{c,s}$,
$Y^{c,s}\equiv X_{+,-}^{c,s}=X_{-,+}^{c,s}$,
come from the nearest-neighbor interaction which mixes
excited pairs in the different Peierls subbands.

With the help of eq.~(\ref{C1}), $\lambda_{+,+}=-\lambda_{-,-}$,
and $\lambda_{+,-}=\lambda_{-,+}$ we can immediately write
\begin{eqnarray}
\langle \hat{\jmath}_{\omega >0}\rangle(V)
&=&  
\langle \hat{\jmath}_{\omega >0}\rangle(V=0)
\nonumber
\\[6pt]
&& - \frac{2V{\cal N}_{\perp}}{La} \sum_{{|q| \leq \pi/(2a)} \atop
{|k| \leq \pi/(2a)} }
\left( {\cos(ka) \atop \sin(ka)} \right)
\Biggl\{
\lambda_{+,+}X^{c,s}
\left( \frac{1}{\omega-E_{+,+}} + \frac{1}{\omega-E_{-,-}} \right)
\\[6pt]
\label{C2}
&&
\phantom{
- \frac{2V{\cal N}_{\perp}}{La} \sum_{|q| \leq \pi/(2a)}
\left( {\cos(ka) \atop \sin(ka)} \right)
\Biggl\{
}
+
\lambda_{+,-}Y^{c,s}
\left( \frac{1}{\omega-E_{+,-}} + \frac{1}{\omega-E_{-,+}} \right)
\Biggr\} \; .
\nonumber 
\end{eqnarray}
To determine~$X^{c,s}$ and $Y^{c,s}$ we have to evaluate
$\hat{V} \hat{u}_{k+q/2,\tau'}^+ \hat{l}_{k-q/2,\tau}^{\phantom{+}}  |0\rangle$.
Since $\hat{V}$ is simple in terms of the original
operators~$\hat{u}_{k}$, $\hat{l}_{k}$ we first have to
apply the inverse transformation of eq.~(\ref{trafoforl}). 
As next step we let~$\hat{V}$ act and then re-transform into
the operators for the Peierls subbands in the last step.

The calculation shows that only four combinations of currents
occur in~$X^{c,s}$ and $Y^{c,s}$, namely,
\begin{mathletters}
\begin{eqnarray}
X^{c,s}(k,q)&=&
- f_{+,+}^*(k,q) J_1^{c,s}(q)
+ f_{+,-}^*(k,q)J_2^{c,s}(q) \\[6pt]
Y^{c,s}(k,q)&=&
f_{+,-}^*(k,q) J_1^{c,s}(q)+
f_{+,+}^*(k,q) J_2^{c,s}(q)
\end{eqnarray}
\end{mathletters}%
with
\begin{mathletters}
\label{C5}
\begin{eqnarray}
J_1^{c,s}(q) &=& \frac{1}{L} \sum_{|p| \leq \pi/(2a)}
\left( {\cos(pa) \atop \sin(pa)}\right)
\left[ f_{+,+}(j_{-,-}-j_{+,+})+f_{+,-}(j_{+,-}+j_{-,+})\right]
\\[6pt]
J_2^{c,s}(q) &=& \frac{1}{L} \sum_{|p| \leq \pi/(2a)}
\left( {\cos(pa) \atop \sin(pa)}\right)
\left[ f_{+,+}(j_{+,-}+j_{-,+})-f_{+,-}(j_{-,-}-j_{+,+})\right] \; .
\end{eqnarray}
\end{mathletters}%
Since the interaction is restricted to nearest-neighbors
the global, i.~e., only $q$-dependent currents~$J_{1,2}^{c,s}(q)$ appear
in the problem.

Equation~(\ref{C2}) becomes
\begin{equation}
\langle \hat{\jmath}_{\omega >0}\rangle(V)
- \langle \hat{\jmath}_{\omega >0}\rangle(V=0)
=
- \frac{2V{\cal N}_{\perp}}{a} \sum_{|q| \leq \pi/(2a) }
\left[ J_1^{c,s}(q)G_1^{c,s}(q)+
J_2^{c,s}(q)G_2^{c,s}(q)\right]
\label{C4}
\end{equation}
with~$G_{1,2}^{c,s}(q)$ given by
\begin{eqnarray}
G_{1,2}^{c,s} (q) &=& \frac{1}{L} \sum_{|k|\leq \pi/(2a)}
\left( {\cos(ka) \atop \sin(ka)}\right)  \biggl[
\left( {-f_{+,+}^* \atop f_{+,-}^*}\right)\lambda_{+,+}
\left(\frac{1}{\omega-E_{-,-}}+\frac{1}{\omega-E_{+,+}}\right)
\nonumber \\[6pt]
&& \phantom{ \frac{1}{L} \sum_{|k|\leq \pi/(2a)}
\left( {\cos(ka) \atop \sin(ka)}\right)  \biggl[  }
+\left( {f_{+,-}^* \atop f_{+,+}^*} \right) \lambda_{+,-}
\left(\frac{1}{\omega-E_{-,+}}+\frac{1}{\omega-E_{+,-}}\right)
\biggr] \; .
\label{capitalG}
\end{eqnarray}
The quantities~$G_{1,2}^{c,s} (q)$ are still operator valued
objects since they contain $\lambda_{\tau,\tau'}$.
Nevertheless these quantities are known. We insert eq.~(\ref{thelambdas})
and use the fact that $|f_{+,+}(-k,q)|^2=|f_{+,+}(k,q)|^2$,
$|f_{+,-}(-k,q)|^2=|f_{+,-}(k,q)|^2$,
and $f_{+,+}^*(-k,q)f_{+,-}^{\phantom{*}}(-k,q)=
f_{+,+}^*(k,q)f_{+,-}^{\phantom{*}}(k,q)$
to show that $G_1^s(q)=G_2^c(q)=0$.
We set $G_1^c(q)\equiv G_1(q)$,
$G_2^s(q)\equiv G_2(q)$, $J_1^c(q)\equiv J_1(q)$,
$J_2^s(q)\equiv J_2(q)$.
They can be expressed in terms of the functions~$F_{1,2,3}(q)$
of eq.~(\ref{capitalF}) as
\begin{mathletters}
\label{CapitalGred}
\begin{eqnarray}
G_1(q)&=&
itea \left[ \hat{x}_q^+ F_1(q) - \hat{x}_{q+\pi/a}^+ F_3(q)\right]\\[6pt]
G_2(q)&=&
itea \left[ -\hat{x}_q^+ F_3(q) + \hat{x}_{q+\pi/a}^+ F_2(q)\right]\; .
\end{eqnarray}
\end{mathletters}%

It remains to determine~$J_{1,2}(q)$.
They can be obtained from their definitions
in eq.~(\ref{C5}) and the result from the equations of motion, eq.~(\ref{C1}),
\begin{mathletters}
\label{C6}
\begin{eqnarray}
J_{1}(q)+\frac{{\cal A}(\omega)}{c}G_1^{+}(q)
&=&
V \left( - J_1(q)F_1(q) + J_2(q) F_3 (q)\right)
\\[6pt]
J_{2}(q)+ \frac{{\cal A}(\omega)}{c}G_2^{+}(q)
&=&
V\left( - J_2(q) F_2(q) + J_1(q)F_3(q) \right)\; .
\end{eqnarray}
\end{mathletters}%
It is not difficult to invert these equations to obtain the
currents explicitly.
The result for the real part of the optical conductivity becomes
{\arraycolsep=0pt
\begin{eqnarray}
{\rm Re}\{\overline{\sigma(\omega >0, V, \delta,\eta)} \}
&=&
{\rm Re}\{\overline{\sigma(\omega >0,\delta,\eta)} \}
+ \frac{2V{\cal N}_{\perp}}{a\omega}  \label{mostimportant}
\\[6pt]
&&{\rm Im}\Biggl\{ \sum_{|q| \leq \pi/(2a) }
\frac{1}{(1+VF_1)(1+VF_2)-(VF_3)^2}
\biggl[G_1^{\phantom{+}}G_1^+ + G_2^{\phantom{+}}G_2^+
\nonumber
\\[6pt]
&& \phantom{ {\rm Im}\biggl\{ \sum_{|q| \leq \pi/(2a) }   }
+V \left(
G_1^{\phantom{+}}G_1^+F_2 + G_2^{\phantom{+}}G_2^+F_1
+ (G_1^{\phantom{+}}G_2^+ + G_2^{\phantom{+}}G_1^+)F_3
\right)\biggr]\Biggr\} \; .
\nonumber
\end{eqnarray}}

As a last step we have to factorize the products over the
functions~$G_{1,2}(q)$.
For the Harris-Lange model we use eq.~(\ref{averagexq}).
With the help of eq.~(\ref{CapitalGred}) it is not difficult
to derive the final result for the average optical conductivity
in the Harris-Lange model, eq.~(\ref{thefinalresultHL}).

\end{appendix}

\newpage
\begin{center} {\bf REFERENCES} \end{center}
\begin{itemize}
\item M.~Abramovitz and I.~A.~Stegun, {\sl Handbook of Mathematical Functions},
(9th~printing, Dover Publications, New York, (1970)).
\item L.~Alc\'{a}cer, and A.~Brau and J.-P.~Farges,
in: {\sl Organic Conductors}, ed. by J.-P.~Farges (Marcel Dekker, New York,
(1994)).
\item A.~A.~Aligia and L.~Arrachea, Phys.~Rev.~Lett.~{\bf 73}, 2240 (1994).
\item N.~Andrei, {\sl Integrable models in condensed matter physics},
in {\sl Proceedings of the summer school on field theoretical methods
(24~August -- 4~September 1992)}, ed.\ by Yu~Lu, (World Scientific, Singapore,
(1995)).
\item D.~Baeriswyl, P.~Horsch, and K.~Maki, 
Phys.~Rev.~Lett.~{\bf 60} (C), 70 (1988).
\item P.-A.~Bares and F.~Gebhard, Europhys.~Lett.~{\bf 29}, 573, (1995).
\item P.-A.~Bares and F.~Gebhard, Journ.\ Low Temp.~Phys.~{\bf 99}, 565 (1995).
\item P.-A.~Bares and F.~Gebhard, J.~Phys.~Cond.~Matt.~{\bf 7}, 2285 (1995).
\item G.~Beni, P.~Pincus, and T.~Holstein, Phys.~Rev.~B~{\bf 8}, 312 (1973).
\item J.~Bernasconi, M.~J.~Rice, W.~R.~Schneider, and
S.~Str\"{a}\ss ler, Phys.~Rev.~B~{\bf 12}, 1090 (1975).
\item J.~D.~Bjorken and S.~D.~Drell, {\sl Relativistic Quantum Mechanics}, 
(McGraw-Hill, New York, (1964)).
\item J.~de Boer, V.~E.~Korepin, and A.~Schad\-schneider,
Phys.~Rev.~Lett.~{\bf 74}, 789 (1995).
\item D.~K.~Campbell, J.~T.~Gammel, and E.~Y.~Loh,
Phys.~Rev.~B~{\bf 38}, 12043 (1988).
\item D.~K.~Campbell, J.~T.~Gammel, and E.~Y.~Loh,
Int.~J.~Mod.~Phys.~B~{\bf 3}, 2131 (1989).
\item D.~K.~Campbell, J.~T.~Gammel, and E.~Y.~Loh, 
in: {\sl Interacting Electrons in Reduced Dimensions},
ed.~by D.~Baeriswyl and D.~K.~Campbell (NATO ASI Series~B~{\bf 213},
Plenum Press, New York, (1989)), p.~171.
\item D.~K.~Campbell, J.~T.~Gammel, and E.~Y.~Loh,
Phys.~Rev.~B~{\bf 42}, 475 (1990).
\item P.~G.~J.~van Dongen, Phys.~Rev.~B~{\bf 49}, 7904 (1994).
\item P.~G.~J.~van Dongen, Phys.~Rev.~B~{\bf 50}, 14016 (1994).
\item F.~H.~L.~E\ss ler and V.~E.~Korepin (ed.), 
{\sl Exactly Solvable Models of Strongly Correlated Electrons}, 
(World Scientific, Singapore, (1994)).
\item J.-P. Farges (ed.), {\sl Organic Conductors}, (Marcel Dekker, New York,
(1994)).
\item J.~L.~Fave, in: {\sl Electronic Properties of Polymers},
ed.~by H.~Kuzmany, M.~Mehring, and S.~Roth,
(Springer Series in Solid State Sciences~{\bf 107}, Springer, Berlin (1992)).
\item H.~Frahm and V.~E.~Korepin, Phys.~Rev.~B~{\bf 42}, 10553 (1990).
\item H.~Frahm and V.~E.~Korepin, Phys.~Rev.~B~{\bf 43}, 5653 (1991).
\item A.~Fritsch and L.~Ducasse, J.~Physique~I~{\bf 1}, 855 (1991).
\item R.~M.~Fye, M.~J.~Martins, D.~J.~Scalapino, J.~Wagner, and W.~Hanke,
Phys.~Rev.~B~{\bf 45}, 7311 (1992).
\item J.-P.~Galinar, J.~Phys.~C~{\bf 12}, L335 (1979).
\item J.~T.~Gammel and D.~K.~Campbell, Phys.~Rev.~Lett.~{\bf 60} (C), 
71 (1988).
\item F.~Gebhard, K. Bott, M.~Scheidler, P.~Thomas, and S.~W.~Koch, 
previous article, referred to as~I.
\item F.~Gebhard, K. Bott, M.~Scheidler, P.~Thomas, and S.~W.~Koch, 
forthcoming article, referred to as~III.
\item F.~Gebhard, A.~Girndt, and A.~E.~Ruckenstein, 
Phys.~Rev.~B~{\bf 49}, 10926 (1994)
\item F.~Gebhard and A.~E.~Ruckenstein, Phys.~Rev.~Lett.~{\bf 68}, 244 (1992).
\item I.~S.~Gradshteyhn and I.~M.~Ryzhik, {\sl A Table of Integrals, Series, 
and Products}, (Academic Press, New York, (1980)).
\item D.~Guo, S.~Mazumdar, S.~N.~Dixit, F.~Kajzar, F.~Jarka, Y.~Kawabe,
and N.~Peyghambarian, Phys.~Rev.~B.~{\bf 48}, 1433 (1993).
\item F.~Guo, D.~Guo, and S.~Mazumdar, Phys.~Rev.~B.~{\bf 49}, 10102 (1994).
\item F.~D.~M.~Haldane, Phys.~Rev.~Lett.~{\bf 67}, 937 (1991).
\item E.~R.~Hansen, {\sl A Table of Series and Products},
(Prentice-Hall, Englewood Cliffs, (1975)).
\item A.~B.~Harris and R.~V.~Lange, Phys.~Rev.~{\bf 157}, 295 (1967).
\item H.~Haug and S.~W.~Koch, {\sl Quantum Theory of the Optical and 
Electronic Properties of Semiconductors}, 
(World Scientific, Singapore, (1990)).
\item J.~Hubbard, Proc.~R.~Soc. London, Ser.~A~{\bf 276}, 238 (1963).
\item N.~Kawakami and S.-K.~Yang, Phys.~Lett.~{\bf A~148}, 359 (1990).
\item N.~Kawakami and S.-K.~Yang, Phys.\ Rev.\ Lett.~{\bf 65}, 3063 (1990).
\item N.~Kawakami and S.-K.~Yang, Phys.~Rev.~B~{\bf 44}, 7844 (1991).
\item S.~Kivelson, W.-P.~Su, J.~R.~Schrieffer, and A.~J.~Heeger,
Phys.~Rev.~Lett.~{\bf 58}, 1899 (1987).
\item S.~Kivelson, W.-P.~Su, J.~R.~Schrieffer, and A.~J.~Heeger,
Phys.~Rev.~Lett.~{\bf 60} (C), 72 (1988).
\item D.~J.~Klein, Phys.~Rev.~B~{\bf 8}, 3452 (1973). 
\item W.~Kohn, Phys.~Rev.~{\bf 133}, A171 (1964).
\item E.~H.~Lieb and F.~Y.~Wu, Phys.~Rev.~Lett.~{\bf 20}, 1445 (1968).
\item S.~K.~Lyo and J.-P.~Galinar, J.~Phys.~C~{\bf 10}, 1693 (1977).
\item S.~K.~Lyo, Phys.~Rev.~B~{\bf 18}, 1854 (1978).
\item G.~D.~Mahan, {\sl Many-Particle Physics}, (2nd~edition, Plenum
Press, New York (1990)).
\item P.~F.~Maldague, Phys.~Rev.~B~{\bf 16}, 2437 (1977).
\item S.~Mazumdar and S.~N.~Dixit, Phys.~Rev.~B~{\bf 34}, 3683 (1986).
\item S.~Mazumdar and Z.~G.~Soos, Phys.~Rev.~B~{\bf 23}, 2810 (1981).
\item F.~Mila, Phys.~Rev.~B~{\bf 52}, 4788 (1995).
\item M.~Ogata and H.~Shiba, Phys.~Rev.~B~{\bf 41}, 2326 (1990).
\item A.~A.~Ovchinnicov, Zh.~Eksp.~Teor.~Fiz.~{\bf 57}, 2137
(1969) (Sov.~Phys.\ JETP~{\bf 30}, 1160 (1970)).
\item A.~Painelli and A.~Girlando, Synth.~Met.~{\bf 27}, A15 (1988).
\item A.~Painelli and A.~Girlando, in: {\sl Interacting Electrons in 
Reduced Dimensions}, ed.~by D.~Baeriswyl and D.~K.~Campbell,
(NATO ASI Series~B~{\bf 213}, Plenum Press, New York, (1989)), p.~165.
\item A.~Painelli and A.~Girlando, Phys.~Rev.~B~{\bf 39}, 2830 (1989).
\item A.~Parola and S.~Sorella, Phys.~Rev.~Lett.~{\bf 60}, 1831 (1990).
\item L.~Salem, {\sl Molecular Orbital Theory of Conjugated Systems}, 
(Benjamin, London, (1966)).
\item A.~Schad\-schneider, Phys.~Rev.~B~{\bf 51}, 10386 (1995).
\item H.~J.~Schulz, Phys.~Rev.~Lett.~{\bf 64}, 2831 (1990).
\item H.~J.~Schulz, in: {\sl Strongly Correlated Electron Systems~II}, 
(Proceedings of the Adriatico Research Conference and Miniworkshop, Trieste, 
Italy, 18~June -- 27~July~1990), ed.~by G.~Baskaran, A.~E.~Ruckenstein, 
E.~Tosatti, and Y.~Lu, (World Scientific, Singapore, (1991)).
\item B.~S.~Shastry, S.~S.~Jha, and V.~Singh (ed.),
{\sl Exactly Solvable Problems in Condensed Matter and Relativistic 
Field Theory}, (Lecture Notes in Physics~{\bf 242}, Springer, Berlin, (1985)).
\item B.~S.~Shastry and B.~Sutherland, Phys.~Rev.~Lett.~{\bf 65}, 243 (1990).
\item W.~T.~Simpson, J.~Am.~Chem.~Soc.~{\bf 73}, 5363 (1951).
\item W.~T.~Simpson, J.~Am.~Chem.~Soc.~{\bf 77}, 6164 (1955).
\item Z.~G.~Soos and S.~Ramesesha, Phys.~Rev.~B~{\bf 29}, 5410 (1984).
\item C.~A.~Stafford and A.~J.~Millis, Phys.~Rev.~B~{\bf 48}, 1409 (1993).
\item C.~A.~Stafford, A.~J.~Millis, and B.~S.~Shastry, 
Phys.~Rev.~B~{\bf 43}, 13660 (1991).
\item P.~Tavan and K.~Schulten, J.~Chem.~Phys.~{\bf 85}, 6602 (1986).
\item C.-Q.~Wu, X.~Sun, and K.~Nasu, Phys.~Rev.~Lett.~{\bf 59}, 831 (1987).

\end{itemize}

\typeout{figure captions}

\begin{figure}[th]
\caption{Reduced average optical conductivity,
$\overline{\sigma_{\rm red}(\omega >0,V)}$,
in the extended Harris-Lange model for $U=2W$
and $V=0, W/2, W$.
A broadening of~$\gamma=0.01W$ has been included.}
\label{hl01}
\end{figure}

\begin{figure}[th]
\caption{Reduced average optical conductivity,
$\overline{\sigma_{\rm red}(\omega >0,V,\delta,\eta)}$,
in the extended dimerized Harris-Lange model for $U=2W$,
$\delta=0.2$, $\eta=-0.06$, $V=0,W/2,W$.
A broadening of~$\gamma=0.01W$ has been included.}
\label{hl11}
\end{figure}

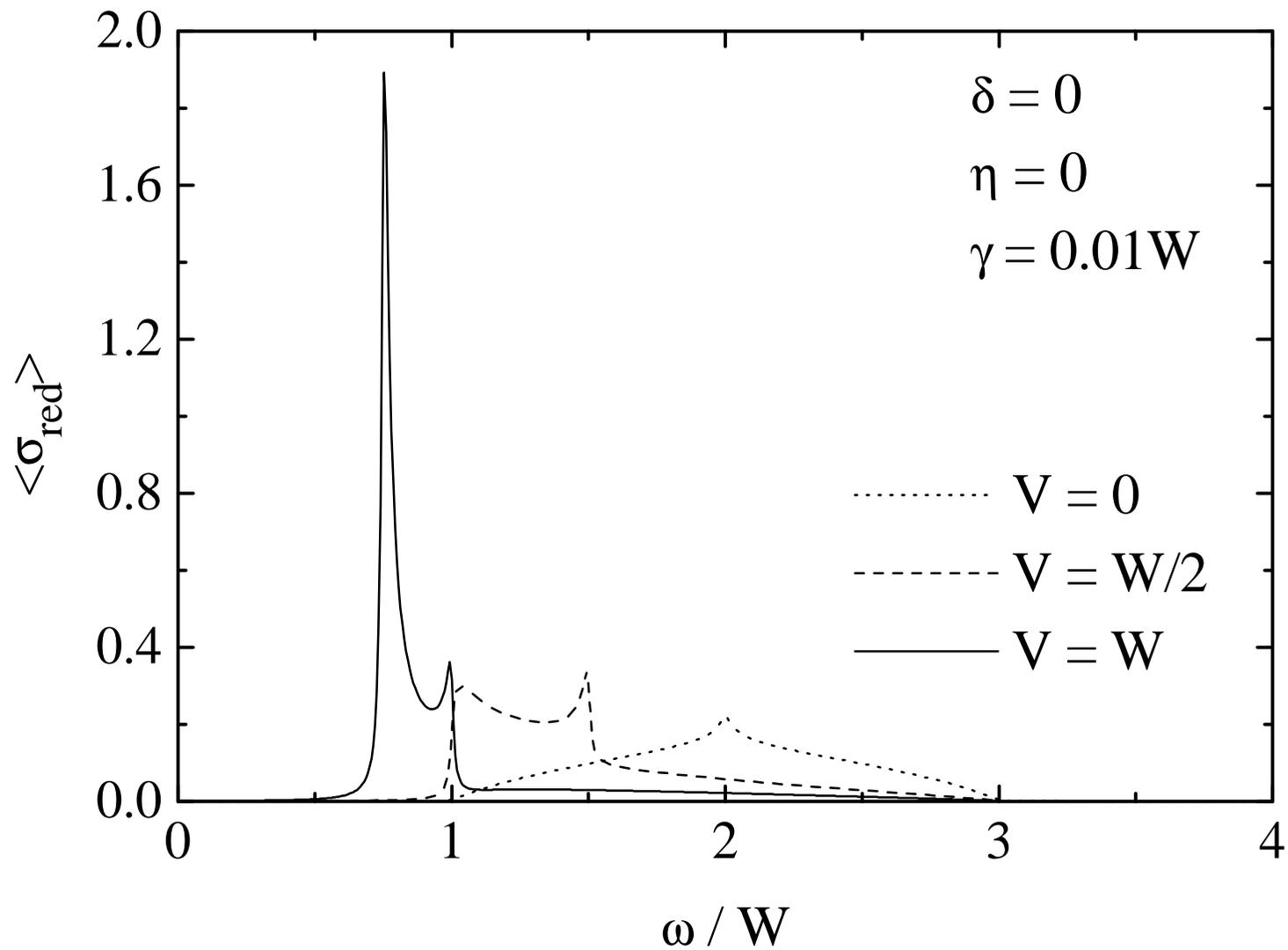
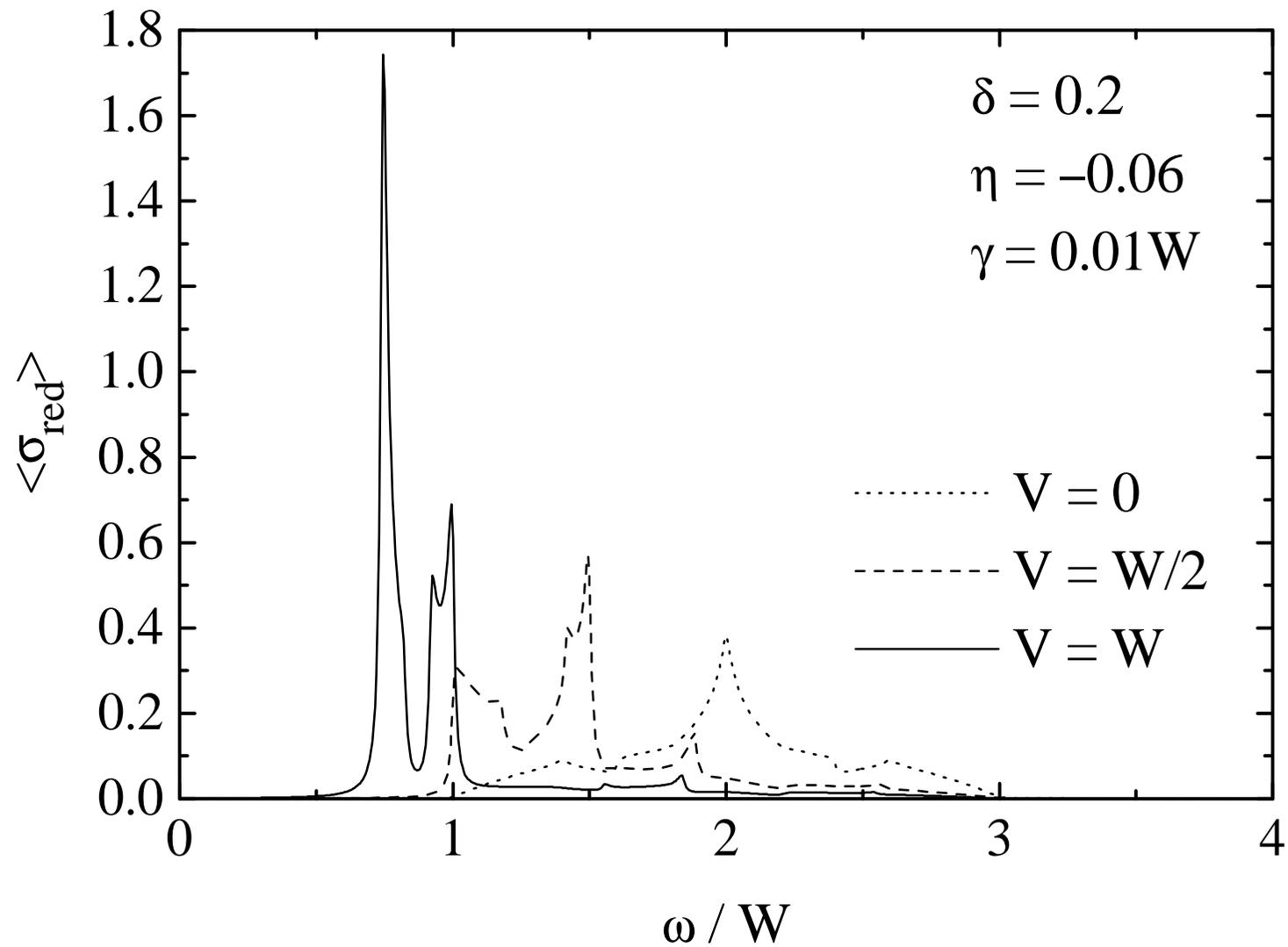
\end{document}